\documentclass[manuscript]{aastex}
\shorttitle{Atomic data and models for \ion{Fe}{2}}
\shortauthors{Bautista et~al.}
\usepackage{enumerate}

\begin{document}
\title{Atomic data and spectral model for \ion{Fe}{2}}

\author{Manuel A. Bautista, Vanessa Fivet\footnote{At present at Astrophysique et Spectrocopie, Universit\'e de Mons - UMONS
B-7000 Mons, Belgium}}
\affil{Department of Physics, Western Michigan University,
Kalamazoo, MI 49008, USA}
\email{manuel.bautista@wmich.edu}

\author{Connor Ballance}
\affil{Department of Physics, Auburn University, Auburn, Alabama 36849, USA}

\author{Pascal Quinet}
\affil{Astrophysique et Spectrocopie, Universit\'e de Mons UMONS,   
B-7000 Mons, Belgium}

\author{Gary Ferland}
\affil{Department of Physics and Astronomy, The University of Kentucky, Lexinton, KY 40506}

\author{Claudio Mendoza}
\affil{Centro de F\'isica, Instituto Venezolano de Investigaciones Cient\'ificas (IVIC), Caracas 1020, Venezuela}

\author{Timothy R. Kallman}
\affil{Code 662, NASA Goddard Space Flight Center, Greenbelt, MD 20771, USA}

\begin{abstract}
  We present extensive calculations of radiative transition rates and electron impact collision strengths for \ion{Fe}{2}. The data sets involve 52 levels from the $3d\,^7$, $3d\,^64s$, and $3d\,^54s^2$ configurations. Computations of $A$-values are carried out with a combination of state-of-the-art multiconfiguration approaches, namely the relativistic Hartree--Fock, Thomas--Fermi--Dirac potential, and Dirac--Fock methods; while the 
$R$-matrix plus intermediate coupling frame transformation, Breit--Pauli $R$-matrix and Dirac $R$-matrix packages are used to obtain collision strengths. 
We examine the advantages and shortcomings of each of these methods, and estimate rate uncertainties from the resulting data dispersion. We proceed to 
construct excitation balance spectral models, and compare the predictions from each data set with observed spectra from various astronomical objects. 
We are thus able to establish benchmarks in the spectral modeling of [\ion{Fe}{2}] emission in the IR and optical regions as well as in the UV \ion{Fe}{2} absorption spectra. Finally, we provide diagnostic line ratios and line emissivities for emission spectroscopy as well as column densities for absorption spectroscopy. All atomic data and models are available online and through the AtomPy atomic data curation environment.
\end{abstract}

\keywords{atomic data -- quasars: absorption lines -- quasars: individual (QSO~2359--1241) -- ISM: individual objects (HH~202) -- ISM: individual objects (Orion) -- 
ISM: individual objects (ESO-H$\alpha$~574,~Par-Lup~3-4) 
-– ISM: jets and outflows –- ISM: lines and
bands --€" stars: pre-main sequence}

%\objectname{J2359-1241}

\normalsize

\section{Introduction}

Reliable quantitative spectral modeling of singly ionized iron (\ion{Fe}{2}) is of paramount astrophysical importance since various fundamental research lines---e.g. active galactic nuclei (AGN), cosmological supernova light curves, solar and late-type-star atmospheres, and gamma-ray-burst (GRB) afterglows---depend on such models. This ion has gained even more attention in recent years with the advent of large-scale observational surveys as well as deeper and high-resolution spectroscopy.

\ion{Fe}{2} spectral modeling firstly requires the detailed treatment of electron impact excitation of metastable levels followed by spontaneous decay through dipole forbidden transitions. The computation of accurate electron impact collision strengths and $A$-values has proven to be cumbersome despite many efforts over several decades. 
The difficulty in describing the \ion{Fe}{2} system arises from the complexity of the effective potential acting on the $4s$, $3d$, and $4p$ electrons. In practice, the wave function of the atomic system is approximated by an anti-symmetrized product of one-electron radial functions determined from the effective potential
\begin{equation}
\Psi(nl\bar r) = V(nl \bar r) + {l(l+1)\over {2r^2}},
\end{equation}
where $V(nl \bar r)$ is the electrostatic potential arising from the nucleus and the $(N-1)$ electrons of the ion, the second term being the centrifugal energy for an electron with orbital angular momentum quantum number $l$. For an effective potential with asymptotic form $-2/r$ the centrifugal term takes the form of a positive barrier for $l\ge 2$, and the effective potential becomes a two-well potential \citep{kar81}. The two potential wells are very different from each other, the inner well is determined by many-electron effects while the outer is mostly hydrogenic. Therefore, slight variations in the potential morphology can lead to large changes in electron localization. For this reason, the atomic structure is very sensitive to orbital relaxation in electron excitation, where the magnitude of such effects varies among the different terms of a given configuration. Furthermore, finding numerical solutions that simultaneously reproduce all the the wave-function conditions
can be difficult and  iterative self-consistent treatments, e.g. Hartree--Fock, may fail to converge or yield poor quality results when compared with measurements (level energies and oscillator strengths). 
Calculations that approximate the wave functions by configuration mixing tend to become intractable, because the collapse of the electron localizations into narrow potential wells gives rise to strong electron exchange interactions. 
Computations that employ distinct non-orthogonal orbitals for each configuration or for each LS term are very difficult 
due to the large number of orbitals that need to be optimized. 
Additional complications arise from spin--orbit coupling and relativistic corrections whose effects on calculated energy levels are comparable to the energy level separations.

There is mounting observational evidence that current [\ion{Fe}{2}] spectral models remain of insufficient accuracy. For example, the predicted [\ion{Fe}{2}] line intensities in the Orion nebula, the archetypical \ion{H}{2} region, disagree with observations by up to several factors \citep[see][]{ver00}. In the context of extragalactic astronomy, there have been significant efforts to use the \ion{Fe}{2}/\ion{Mg}{2} emission ratio as a direct Fe/Mg abundance indicator in quasars. According to models of
cosmological nucleosynthesis, Fe enrichment trails behind $\alpha$-element enrichment until $\sim 1{-}2$~Gyr after the initial star-formation burst; consequently, many groups are actively 
trying to find such a point of inflexion \citep[e.g.][]{kur07, sam09}. However, most of these efforts have been inconclusive due to the large scatter in the Fe/Mg ratio. Uncertainties arise from the use of \ion{Fe}{2}(UV)/\ion{Mg}{2} as an abundance indicator, which are exacerbated by the fact that the classical photoionization models fail to account for the \ion{Fe}{2}(${\lambda}4570)$/\ion{Fe}{2}(UV) ratio by an order of magnitude; therefore, \ion{Fe}{2} abundance estimates derived from using current spectral models are unreliable \citep[see][]{col80, col00, bal04}.

We report new calculations of $A$-values and collision strengths for the lowest 52 even-parity levels of \ion{Fe}{2}. The atomic data were computed using a multi-platform approach, where most of the state-of-the-art numerical methods of atomic physics have been used in a concerted effort to provide consistency checks and comparisons. Furthermore, we present NLTE spectral models whose predictions are benchmarked with the available astronomical spectra. These comparisons provide stringent tests on the quality of the atomic data.

We also present a detailed analysis of the inherent uncertainties in the atomic data and their implications in NLTE spectral models; for this analysis, we follow the method described by \cite{bau13}. Under steady-state balance the population of a level $i$ is given by
\begin{equation}
N_i={\sum_{k\ne i} N_k(n_e q_{ki} + A_{ki})\over
n_e \tau_i \sum_{j\ne i} q_{ij} + 1} \tau_i
=
{\sum_{k\ne i} N_k(n_e q_{ki}\tau_i + b_{ki})\over
n_e \tau_i \sum_{j\ne i} q_{ij} + 1},
\end{equation}
where $n_e$ is the electron density, $A_{ki}$ is the Einstein spontaneous radiative decay rate from level $k$ to level $i$,
$b_{ki}$ is the branching ratio, $q_{ki}$ is the electron impact transition rate coefficient, and
$\tau_i=(\sum_{j<i} A_{ij})^{-1}$ is the level lifetime. Expressing level populations in terms of lifetimes and branching ratios, as opposed to using only $A$-values, has the practical advantage that lifetimes are generally dominated by a few strong transitions; i.e. they are generally more accurate than individual rates. Therefore, this level-population formalism gives a more clear insight into the propagation of atomic data uncertainties. 

\section{Atomic structure and radiative calculations}

Realistic representations of the atomic structure are needed to obtain accurate energy levels, line wavelengths, and $A$-values, and also because such representations are the basis of reliable scattering calculations. For this work we use a combination of numerical methods: the pseudo-relativistic Hartree--Fock ({\sc hfr}) code of \citet{cow81}; the Multiconfiguration Dirac--Fock ({\sc mcdf}) code \citep{dya89}, and the scaled Thomas--Fermi--Dirac central-field potential as implemented in {\sc autostructure} \citep{bad97}.

\subsection{{\sc hfr} calculations}

{\sc hfr} uses a superposition of configurations approach to account for configuration interactions. The code solves the Hartree--Fock equations 
for each electronic configuration. Relativistic corrections are also included in this set of equations. The radial parts of the multi-electron Hamiltonian can be adjusted empirically to reproduce the spectroscopic energy levels in a least-squares fit procedure. These semi-empirical corrections are used to account for the contributions from higher order correlations in the atomic state functions. 

The following configurations were explicitly included in the physical model: $3d\,^64s$, $3d\,^7$, $3d\,^54s^2$, $3d\,^65s$, $3d\,^64d$, $3d\,^65d$, $3d\,^54p^2$, $3d\,^54d\,^2$, $3d\,^54s4d$, $3s3p^63d\,^74s$, $3s3p^63d\,^8$, and $3s3p^63d\,^64s^2$. This configuration expansion extends the one used in the previous {\sc hfr} calculation by \citet{qui96} by including $3d\,^54d\,^2$ and $3s3p^63d\,^64s^2$. In order to minimize the discrepancies between computed and experimental energy levels, the {\sc hfr} technique was used in combination with a well-known least-squares optimization of the radial parameters. The fitting procedure was applied to $3d\,^64s$, $3d\,^7$, and $3d\,^54s^2$ with the experimental energy levels compiled by \citet{sug85}. In the absence of configuration interaction, the $3d\,^7$ and $3d\,^54s^2$ configurations are described by four parameters, namely the average energy $E({\rm av})$, the Slater integrals $F^2(3d,3d)$ and $F^4(3d,3d)$, and the spin--orbit parameter 
$\zeta(3d)$, while for $3d\,^64s$ the exchange interaction integral $G^2(3d,4s)$ is also required. In addition to these parameters, effective interaction parameters such as $\alpha$ and $\beta$, associated with the excitation out of the $3s$ and $3p$ subshells into the $3d$, are used in the fit. 
The average deviation between computed and experimental levels was found to be equal to 78~cm$^{-1}$.

\subsection{{\sc autostructure} calculations}

{\sc autostructure} \citep{bad97,bad11} computes CI state wave functions built using single-electron orbitals generated from a scaled Thomas--Fermi--Dirac--Amaldi (TFDA) potential. The scaling factors for each orbital are optimized in a multiconfiguration variational procedure minimizing a weighted average of $LS$ non-relativistic term energies or term energies including the effects of one-body Breit--Pauli (BP) effects. Spin--orbit coupling and BP operators are introduced as perturbations to obtain fine-structure relativistic corrections. Semi-empirical corrections can also be applied to the multi-electron Hamiltonian, where the theoretical $LS$ term energies are corrected in order to reproduce the centers of gravity of the available experimental multiplets.

 \citet{bau08} introduced non-spherical multipole corrections to the TFDA potential to account for some of the electron correlation effects. This work also considered alternative optimization techniques of the scaling parameters for systems where the spin--orbit and relativistic effects are important. For the present work, we found necessary to use such developments but with a single variation. The modified potential is written as
\begin{equation}
V^c(r ) = V^{TFDA}(r;\lambda ) +\lambda_d {8\over{3\pi}}\left[{1\over{s^2}}\int_0^s \rho(r_2)s_2^3 ds_2
+s \int_s^{s_0} \rho(s_2) ds_2 \right ]
\end{equation}\label{NewTFDAc}
where $s=r/\lambda_{dr}$, and $\rho(r )$ is the electron charge density given by
\begin{equation}
\rho(r ) = {1\over 2\pi^2}\left\{{1\over \pi} + \left[{1\over \pi^2}+V_0-V(r )\right]^{1/2}\right\}
\end{equation}
with
\begin{equation}
V_0 = -\frac{15}{16\pi^2}-{2(Z-N)\over r_0},
\end{equation}
$Z$ being the nuclear charge and $N$ the electron number of the system. This potential is different from the original in \citet{bau08} inasmuch as being limited to a single dipole correction modulated by the scaling parameter $\lambda_d$, but more importantly, the radial dependence of this dipole correction is now scaled by an additional parameter $\lambda_{dr}$.

We have performed numerous calculations with different configuration expansions, starting with those in previous work then adding more configurations and/or using corrected TFDA potentials. From these computations we have arrived at the following general conclusions:
\begin{enumerate}[(a)]
  \item The inclusion of the $3s^23p^63d\,^6\overline{4d}$ configuration with a $\overline{4d}$ pseudo-orbital is essential to obtain a satisfactory structure for the low even-parity levels.
       
  \item The $3s^23p^63d\,^6nl$ configurations with $n>4$ have little effect on the structure of the low even-parity levels.
  
  \item Very large configuration expansions tend to give inferior results than well-selected, concise representations.
   
  \item The $A$-values for forbidden transitions among the low even-parity levels are very sensitive to the predicted energy of the $3d\,^54s^2\ ^6S$ term. 
      
  \item It is difficult to reproduce the observed energy difference between the ground state $3d\,^64s\ ^6D$ and the first excited term $3d\,^7\ ^4F$ using the standard TFDA potential.
      
  \item The magnitude of the spin--orbit correction to the energy separation between the $3d\,^64s\ ^6D$ and $3d\,^7\ ^4F$ states is large; therefore, an optimization of the atomic orbitals based on $LS$ energies can be misleading.
\end{enumerate}

Table~\ref{table:AUTOconfigs} shows a selected set of calculations in the present study. The spectroscopic configurations $3d\,^64s$, $3d\,^7$, and $3d\,^54s^2$ are common to all expansions; $3d\,^6\overline{4d}$ is also common to all as it accounts for essential relaxation effects in configurations involving the $3d$ orbital. The first listed calculation is labeled ``BP extend TFDAc", and is built up from the original expansion of \cite{bau96} but taking into account additional configurations. The second calculation (``Q96+$4d\,^2$") includes the same expansion as our {\sc hfr} calculation. In ``7-config", we tried to use the smallest expansion possible, comparable to the six configuration expansion created with the {\sc mcdf} method (see next section), but with the addition of a $\overline{4d}$ orbital, which could not be optimized in the MCDF method. While this expansion seems very small it was surprising to find that it yields some of the A-values that best agree with observed spectra (see Section 3). The last calculation, ``BP new-TFDAc", is similar to ``BP extend TFDAc" but makes use of the new correlated TFDA potential quoted in Equation~3.

Table~\ref{table:LSenergies} lists the term energies for the various calculations, showing the non-relativistic energies calculated in $LS$-coupling as well as the term-averaged BP energies. As a reference, we start by looking at the \ion{Fe}{2} expansion of \citet{bau96}. This model optimization scheme was based on the non-relativistic $LS$ energies which are seen to be underestimated by 33\% for the $3d\,^7\ ^4F$ state and overestimated by 23\% for $3d\,^64s\ ^4D$. It correctly predicts the relative order of the first six $LS$ terms, permutes the order of the seventh and eighth terms ($3d\,^7\ ^2H$ and $^2D$, respectively) and mis-assigns the positions of most of the higher terms. The important $3d\,^54s^2\ ^6S$ term is predicted to lie tenth relative the ground term in contrast to the observed position (12th). By including relativistic and spin--orbit coupling effects in this model, the predictions change considerably and for the worse. The term-averaged energy for the $3d\,^7\ ^4F$ state is now overestimated by a factor of 2.4, and all the other terms within the $3d\,^7$ configuration also deviate farther from the observed energies. The relative order of the energy terms also deteriorates by including spin--orbit effects; for instance, the
$3d\,^54s^2\ ^6S$ term is predicted to be as low as the eighth position. These large changes in the atomic structure of this model, caused by spin--orbit effects, are consequences of the limitations in optimizing the \ion{Fe}{2} atomic model on the non-relativistic $LS$ energies as traditionally performed in {\sc autostructure}.

The polarized TFDA potential of \citet{bau08} does not improve by itself the atomic model as evidenced by the predicted energies obtained with ``BP extend TFDAc" (see Table~\ref{table:LSenergies}). The expansion ``Q96+$4d\,^2$-corr" is significantly smaller and is expected to be of inferior quality. This is confirmed with the poor energy prediction for the $3d\,^54s^2\ ^6S$ state as well as the significantly higher core energies. On the other hand, relative to the ground term, this model predicts energies  for the $3d\,^64s$ and $3d\,^7$ that compare with experiment as favorably as the other models. The last two columns in Table~\ref{table:LSenergies} show the results obtained with the new polarized TFDA potential (Equation~3) and an optimization scheme based on the (2$j$+1)-averaged energies. In order to optimize the energy of the first excited $3d\,^7\ 4F$ term to within 3\% with respect to experiment, its non-relativistic $LS$ energy becomes lower than that of the  $3d\,^64s\ ^6D$ ground term. This latter model is also better than the previous models as far as the relativistic core energy, despite the fact that the non-relativistic $LS$ core energy actually looks higher. This finding once again confirms that, in order to optimize an \ion{Fe}{2} wave-function representation, the spin--orbit and relativistic energy corrections must be taken into account.

Recent versions of {\sc autostructure} enable the inclusion of one-body BP relativistic operators in the hamiltonian. Thus, orbitals can be optimized 
on term energies that include these relativistic effects, albeit missing spin-orbit splitting of fine structure levels. We verified that the dominant relativistic corrections to the term energies are indeed accounted for by the one-body operators. Hence, this new feature could have been used to optimize the \ion{Fe}{2} system instead of the approach of Bautista (2008). Though, neither of 
these two techniques was available in the code {\sc superstructure} \cite{eissner} used in previous works.

Table~\ref{table:Elevels} gives a complete list of the energy levels considered in this work, where the assigned level indexes will be the reference for the remaining of the paper.

\subsection{{\sc mcdf} calculations}

Atomic descriptions were obtained within the multiconfiguration Dirac--Fock ({\sc mcdf}) framework with the GRASP0 (General-purpose Relativistic Atomic Structure Package) code \citep{gra80a, gra80b, mck80, nor04}.  Our best models included six and 12 non-relativistic configurations. The six configurations model included 
$3d\,^64s$, $3d\,^7$, $3d\,^54s^2$, $3d\,^64p$, $3p^43d\,^9$, and $3p^63d\,^9$ (with an empty 3s orbital). The twelve configuration model included 
$3d\,^64s$, $3d\,^7$, $3d\,^54s^2$,  $3d\,^64d$, $3p\,^43d\,^9$, $3p^53d\,^64s^2$, $3p^53d\,^8$, $3s3p^63d\,^8$, $3s3p^63d\,^74s$, and $3p^63d\,^84s$. 

We found in the {\sc autostructure} calculations that the inclusion of the $3d\,^64d$ configuration
was important to obtain an accurate representation of the atomic structure. However, we were unable to obtain a fully converged $4d$ orbital by using the EAL (Extended Average Level) optimization of the 63 metastable levels. This option optimized a weighted trace of the Hamiltonian, the weighting factor being proportional to the statistical weights $(2J+1)$ of the levels considered. We therefore optimized all the other orbitals ($1s$, $2s$, $2p$, $3s$, $3p$, $3d$, $4s$, and $4p$) within an EAL optimization procedure, and obtained the $4d$ orbital in a single-configuration {\sc mcdf} calculation.

The predicted energies of the twelve configuration models are slightly better than from the smaller model, yet rather poor in comparison with experimental energies. The six configuration model is much better suited for subsequent scattering calculations (see Section 4).
The average agreement between the experimental and theoretical energy levels is around 20\%. Major discrepancies are observed in levels belonging to the low-lying term $3d\,^64s\ a^4D$ (around 75\%). The relative ordering of the metastable even parity states is also somewhat problematic. When using either one of these models to compute radiative lifetimes it is found that the computed values  disagree with all other calculations and experiments for most levels (up to an order of magnitude for the most sensitive levels) indicating that level mixing may not be properly described.

In an attempt to improve the {\sc mcdf} model, we performed a calculation using a $4d$ orbital from a multiconfiguration {\sc autostructure} calculation. This 
emploied a utility program to read the {\sc autostructure} orbitals from a disk file to transcribe the orbitals from a linear radial mesh to the GRASP0 exponential prescription. However, this procedure neither improved the agreement between the theoretical and experimental energies nor the lifetimes.

\subsection{Lifetimes, branching ratios, and $A$-values}
\label{rad}

As discussed in Section~1, it is convenient in NLTE spectral models to explicitly write level populations in terms of level lifetimes and branching ratios rather than transition probabilities ($A$-values). We discuss the radiative data from the various approximations in terms of these quantities, and compare them with previous results in the literature. From these comparisons, we arrive at recommended values and estimated uncertainties.

Table~\ref{table:lifetimes} tabulates the present level radiative widths ($A_i=\tau_i^{-1} = \sum_j A(i\to j)$) as well as previously published values. The last two columns indicate our recommended values and estimated uncertainties, which are obtained from the mean values and statistical dispersion among all the available atomic data \citep[see][]{bau09, bau13}. This procedure was applied to all levels with a few exceptions that yield grossly discrepant results with respect to the rest of the calculations which are then removed. This was the case of levels 21, 28, 31, 43, 44, 47, and 58. Another exception is the lifetime of level 6 ($3d\,^7\ ^4F_{9/2}$), which is of great importance in the excitation of the \ion{Fe}{2} ion \citep{bau13}. The lifetime for this level is determined by a single transition, $3d\,^7\ ^4F_{9/2}-3d\,^64s\ ^6D_{9/2}$, which is very difficult to render accurately due to configuration-interaction (CI) cancelation effects. The standard $A$-value deviation in this transition is $\sim 80$\%. Given the importance of this transition, it had to be treated with much more detail.

\begin{figure}
  \scalebox{0.7}{\includegraphics{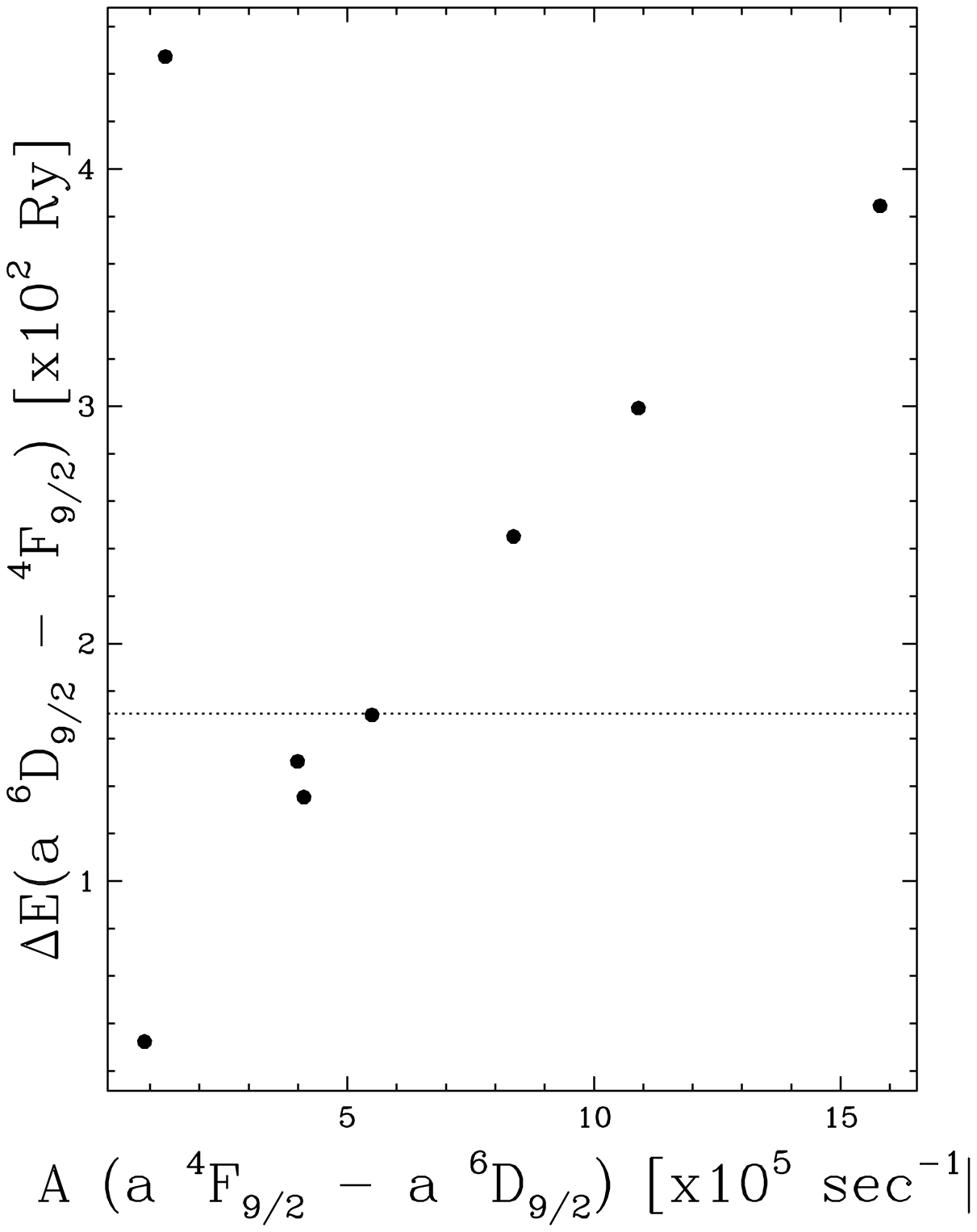}}
  \caption{Theoretical $A$-values as a function of energy separation for the $3d\,^7\ ^4F_{9/2} - 3d\,^64s\ ^6D_{9/2}$ transition calculated with the modified TFDA potential model (Equation 3). The horizontal dotted line depicts the experimentally observed energy separation.}
  \label{fig:AvsE}
\end{figure}

In order to compute the rate for the highly mixed $3d\,^7\ ^4F_{9/2} - 3d\,^64s\ ^6D_{9/2}$ transition, the observed transition energy must be first reproduced, which is only achieved by the model using the newly modified TFDA potential (Equation 3).
In Figure~\ref{fig:AvsE} we plot the calculated $A(3d\,^7\ ^4F_{9/2} - 3d\,^64s\ ^6D_{9/2}$) value {\em vs}. its predicted
transition energy. These values are obtained from the NewTFDAc model by varying the optimization parameters in the potential. The $A$-value predicted by this model is our recommended value in Table~\ref{table:lifetimes}, assigning a conservative uncertainty of 30\%. While we expect these models to give a reasonably reliable A-value for the $3d\,^7\ ^4F_{9/2} - 3d\,^64s\ ^6D_{9/2}$ transition, it seems yield poor results for transitions involving higher excitation multiplets. This is apparent in Table~\ref{table:lifetimes}  by the fact that the NewTFDAc model yields many outliners.

In regards to the overall accuracy of the lifetimes, the observed dispersion between results of different models give an indication of how well converged the results are, thus we suggest that such dispersion can be use as an uncertainty indicator. In this sense, we find that the lifetimes for the lowest 16 levels of \ion{Fe}{2}, which are responsible for the infrared and near-infrared spectra, are known within 10\% or better with only a few exceptions. The lifetimes for higher levels, which yield the optical spectrum have uncertainties that range between $\sim$10 and 30\%.
As to which particular atomic model is the most accurate, that is difficult to say from a purely theoretical point of view, thus further analysis is needed in view of experimental and astronomical spectroscopic information.

In Table~\ref{table:branching} we present a branching-ratio sample from our various computations as well as published radiative
data; average values and standard errors are also shown (the complete table is available in electronic form). It may be appreciated that the stronger transitions, of more practical interest,
 are typically those with large branching ratios, i.e. near unity and carry the lesser uncertainties.

Table \ref{table:branching} illustrates that while branching ratios from the levels of the lowest two multiplets of \ion{Fe}{2} are well known the uncertainties increase for higher levels. For example, in the branching ratios from level 10 there are two models that yield discrepant results. 
The ``BP extend TFDAc"  model yields the largest branching ratios to levels of the a~$^4$F multiplet at the expense of diminishing the ratios to the ground a~$^6$D ground multiplet. This can be understood from the fact that the ``BP extend TFDAc" model yields a very high energy for the a~$^4$F term, thus increasing the overlap of these levels with the a~$^4$D levels. The other discrepant model is the ``Q96+4d$^2$-corr", 
which yields the smallest branching ratios to the a~$^4$F levels. 

\begin{figure}
  \scalebox{0.6}{\includegraphics[angle=-90]{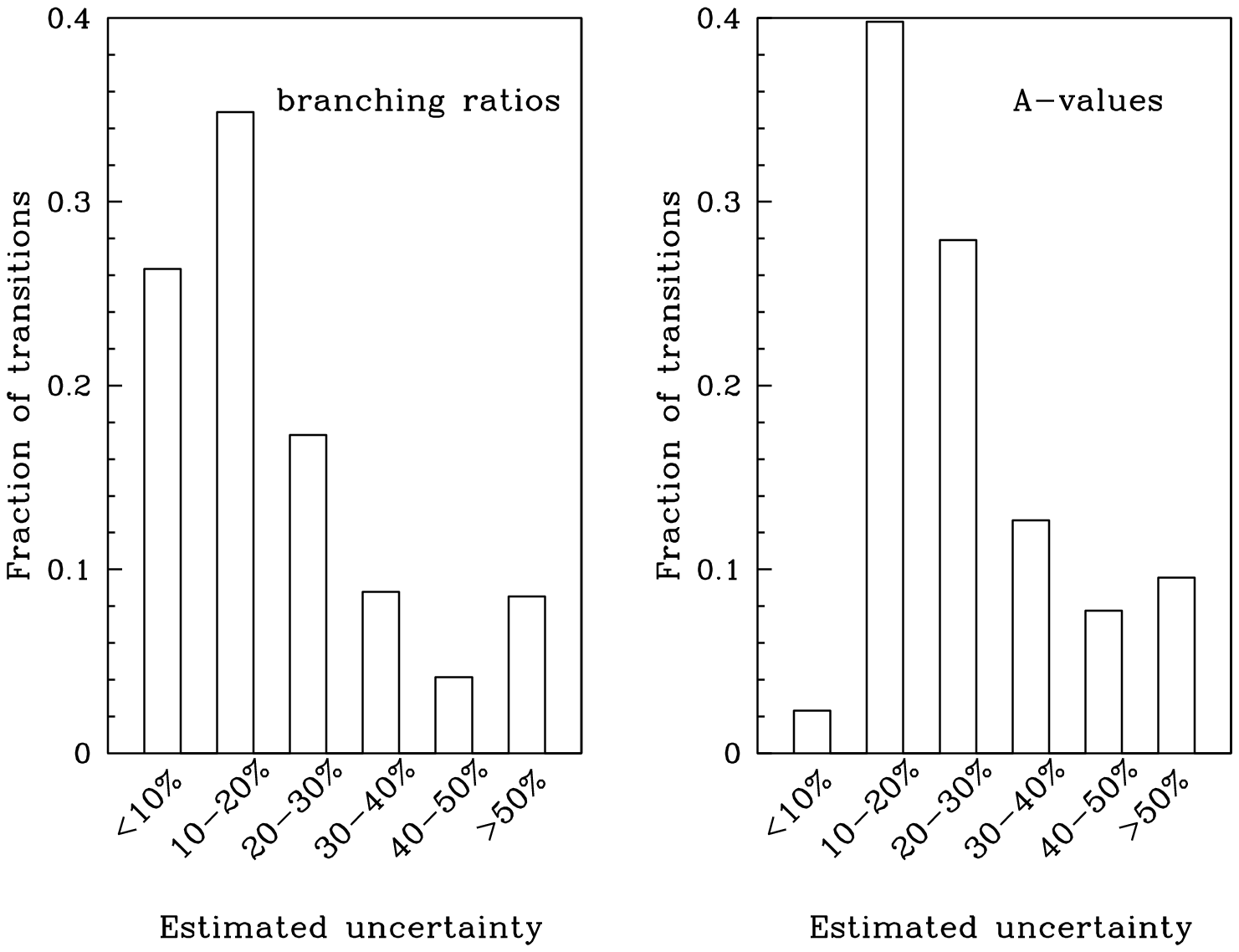}}
  \caption{Branching-ratio and A-values error distributions for dipole-forbidden transitions among metastable levels of \ion{Fe}{2}.}
  \label{fig:brancherr}
\end{figure}

Figure~\ref{fig:brancherr} (left panel) shows the estimated uncertainty distribution of 387 branching ratios greater than 0.01. 
The right panel of this figure depicts the $A$-values uncertainty distribution. It may be seen that more than a quarter of all branching ratios 
are constrained to within 10\% and about two thirds to within 20\%. The uncertainties in the absolute $A$-values can be seen as the combined uncertainties of both branching ratios and lifetimes; therefore, only a small fraction ($\sim$0.02) is determined to better than 10\%. A positive result nonetheless is that over two fifths of all transitions are in accord to within 20\% and nearly three quarters to within 30\%. We expect that the majority of observable lines and transitions regulating the ion population balance are sufficiently accurate.

The complete set of $A$-values for transitions among the levels considered here is available online from AtomPy \citep{men14} or by request from the authors. The present data set should provide a solid platform for the modeling of \ion{Fe}{2}
spectra. Therefore, we argue that further work in improving the quality of the radiative data should concentrate
on specific transitions of observational interest rather than on the whole lot.

\section{Radiative data benchmarks}

In Section~\ref{rad} we presented recommended lifetimes (radiative widths), radiative branching ratios, and $A$-values as a result of a critical assessment of the available data and from the various calculations carried out here. We also provided uncertainty estimates for these quantities based on the data statistical dispersion. In this section we make use of laboratory measurements and astronomical observations to benchmark the quality of the theoretical atomic data, and try to assert the reliability of the recommended data and their estimated uncertainties. It is important to point out here that we have used observed energies in the calculations of all radiative rates. 

The theoretical lifetimes presented in Section~\ref{rad} are compared in Table~\ref{table:lifetcomp} with measurements from the Ferrum Project \citep{har03, gur09}. It is seen that the lifetimes are not reproduced accurately as a whole by any of the calculations. The very long life of the level 43 is particularly difficult to compute accurately, and only the results of Deb and Hibbert (2011) agree with experiment. The results of the ``new TFDA" model exhibit the largest differences with this experiment. This is to be expected since this model was optimized on transitions among the first two terms and is known not to represent well the highly excited levels. 
The scatter in the results of the various calculations owes to the fact that the levels are highly mixed and even small variations in the mixing coefficients in the different representations can lead to cancelation effects in the line strengths. Though, provided that all the dominant configurations are included in all calculations, the computed line strengths should all scatter around the correct answer. Thus, we suggest that 
by averaging over the results of a number of reasonable representations of the ion one should arrive to reliable results. This approach seems to be well supported by Table~\ref{table:lifetcomp}, where our recommended values, resulting from averages of various results, agree with experiments to within the estimated error margins.

We now compare the observed and theoretical intensity ratios between emission lines arising from the same upper level. 
In taken ratios of lines from the same upper level the population of that level, any dependance on the physical conditions, cancels out 
and the ratios are given by 
\begin{equation}
{F(i\to j)\over F(k\to l)} = {A(i\to j)\over A(k\to l)}\times {\lambda(k\to l)\over
\lambda(i\to j)} = {b(i\to j)\over b(k\to l)}\times {\lambda(k\to l)\over
\lambda(i\to j)}\ .
\end{equation}
The advantage of looking at these ratios is that they depend only on $A$-values or, more specifically, on branching ratios regardless of the physical conditions of the plasma. Therefore, the ratios ought to be the same in any source spectra, 
provided the spectra have been corrected for extinction. \ion{Fe}{2} yields the richest spectrum of any astronomically abundant chemical species; thus, its high-resolution optical and near-IR forbidden lines are the best suited for the present evaluation.
137 [\ion{Fe}{2}] lines are found in the HST/STIS archived spectra of the Weigelt blobs of $\eta$~Carinae. 
78 [\ion{Fe}{2}] lines are present in the deep echelle spectrum ($R=30\,000$) of the Herbig--Haro object (HH~202) in the Orion Nebula \citep{mes09}, and 55 [\ion{Fe}{2}] lines have been measured in the X-shooter ($3000{-}25\,000$~\AA) spectrum of the jet of the pre-main-sequence star SEO-H$\alpha$~574 \citep{gia13}.

The $\eta$ Carinae spectra are at $-28^\circ$ position angle (PA), and include Weigelt B and D positioned at $0.15''$
and $0.28''$. 
Six medium dispersion spectra ($R=6000{-}10\,000$) of the blobs have been recorded between 1998 and 2004 at various orbital phases of the star's 5.5-year cycle.
Two complete spectra, recorded during the broad high state and the several month long low state, were offset onto Weigelt D observations at ${\rm PA}=68^\circ$. Additional observations centered on $\eta$ Carinae recorded the spectrum of Weigelt C offset to the southwest. Over 900 HST/STIS spectral segments were recorded of $\eta$ Carinae and the Homunculus
(reduced line-by-line spectra are available online\footnote{http://etacar.umn.edu/ and http://archive.stsci.edu/prepds/etacar/}). To estimate accurate line fluxes we needed to control contamination effects by
stellar radiation, arbitrary continuum placement, and unidentified line blends. The first two issues were mitigated by choosing slit orientations that avoided the bright central star, and by finding spectral extractions that minimized the continuum. To find blends and other extraneous features affecting the lines, we assumed Gaussian profiles defined by the HST/STIS instrument response, and determined average centroid velocity and full-width-at-half-maximum for each species by fitting a sample of the stronger features with no known blends and clean continua. Then, all line fluxes were measured by fitting the features with Gaussians constrained by the parameters determined above, but allowing them to vary within the estimated uncertainties.
We carried out five independent spectral reductions with up to four measurements of every observation for different spectral extractions along the CCD and different assumptions about the continuum and noise levels. Our visible spectrum measurements showed that line centroids and widths can be constrained to within 2~km\,s$^{-1}$; therefore, blends and contamination that affected line peak by more than 0.03~\AA\ could be readily identified. Line fluxes of strong, unblended features were measured with an accuracy of better than 10\%. For other less certain lines, we at least obtained robust error bars.

Measured line fluxes in $\eta$~Carinae must be corrected for extinction, and this affects the comparison between observed and theoretical line ratios. While the extinction curve toward the Weigelt blobs and other regions of the nebula are not well understood, most spectroscopic evidence suggests that the extinction curve is well described by $A_V=2.1$ and $R_V=3.1$
\citep[see][for a discussion]{bau11}. Here we explore the extinction magnitude and selective extinction, and find that the above parameters yield the best agreement with the theoretical line ratios. Moreover, it is found that even large variations in the adopted visual extinction magnitude would not change the our conclusions regarding the \ion{Fe}{2} $A$-values choice.

\begin{figure}
  \scalebox{0.6}{\includegraphics{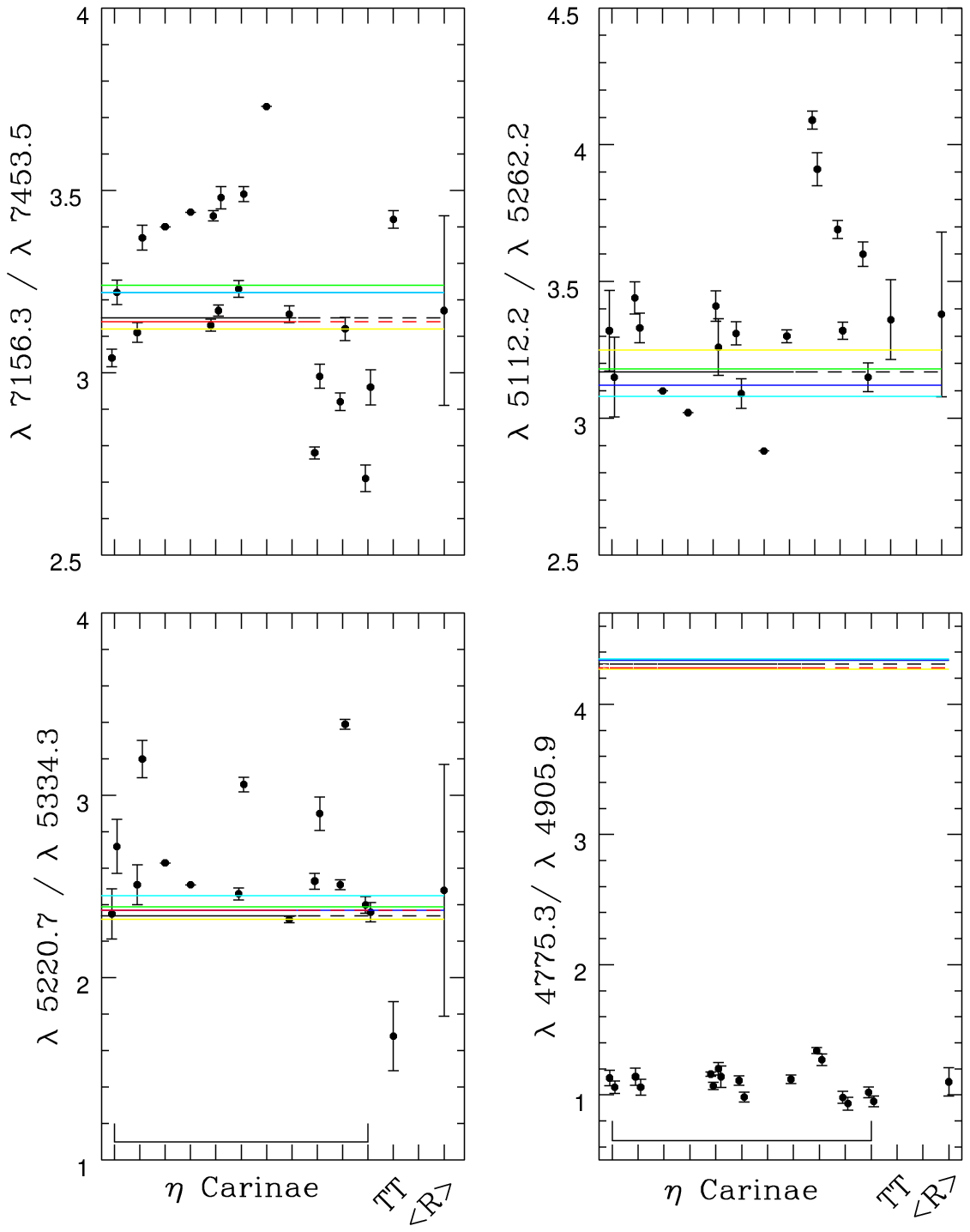}}
  \caption{Emission-line ratios among transitions from the same upper level. The first nine points from left to right ('$\eta$ Carinae') are the results from our measured intensities in the HST/STIS spectra of the Weigelt blobs of $\eta$~Carinae. The tenth point ('TT') is the measured ratio in the X-shooter spectrum of SEO-H$\alpha$~574. The last point to the right ('$<$R$>$') depicts the average of all measurements and uncertainties given by the standard deviation. The horizontal lines represent the predictions from several different computations of $A$-values.}
  \label{fig:obsratios}
\end{figure}

Comparisons of measured line ratios show significant systematic errors that are easily overlooked, and by analyzing multiple measurements of the same line ratio, we attempt to minimize such systematic errors and gain insight on the observational accuracy.
Regarding the $\eta$~Carinae spectra, we find 107 reasonably well-measured line ratios that are defined as
\begin{equation}
{\rm ratio} = \frac{{\rm max}(F1, F2)}{{\rm min}(F1, F2)}\ ,
\end{equation}
where $F1$ and $F2$ are the measured fluxes of two lines from the same upper level. The minimum of the two fluxes is in the denominator such that the line ratios are unconstrained, and equally weighted when compared with the theoretical expectations. Figure~\ref{fig:obsratios} shows a line-ratio sample from various observed spectra as well as from multiple
theoretical determinations, where some of the most common traits are illustrated. A regular feature is the scatter in the observed values which greatly exceeds the estimated individual errors. Also the scatter in the measured line ratios often exceeds that of the theoretical predictions. In some cases (see the bottom-left panel of Figure~\ref{fig:obsratios}), there are systematic differences between the measured ratios in $\eta$~Carinae, SEO-H$\alpha$~574, and HH~202. 
Note that we only measured the lines in $\eta$~Carinae, while for SEO-H$\alpha$~574 and HH~202 we adopt the published line intensities. Thus, the fact
that there are discrepancies between the ratios from the latter two sources is further evidence of unaccounted systematic errors in the measurements, independent of our own work. 
In five cases, the differences
between observation and theory are in excess of the scatter of the different methods (see the bottom-right panel of Figure~\ref{fig:obsratios}). (Plots of all 106 line ratios extracted from observations are available from the authors upon
request.)
The results of these comparisons should serve to warn researchers against the temptation to try to derive atomic parameters from a single set of observed spectra.

The adopted line ratios for the present study are the mean observed values and their uncertainties are given in terms of their standard deviations. These line ratios are then compared with theory and our recommended branching ratios. It is worth mentioning that, while the ratios determined from $A$-values and branching ratios are equivalent, the error margins determined from the latter are smaller and thus preferable. Table~\ref{table:chis} shows reduced $\chi^2$ statistical indexes for this comparison which are defined by the expression
\begin{equation}
\overline{\chi}\,^2 = \sqrt{{1\over N}\sum {(R_{\rm ob}-R_{\rm th})^2\over (\delta R_{\rm ob}^2+\delta R_{\rm th}^2)}},
\end{equation}
where $R_{\rm ob}$ and $R_{\rm th}$ are the observed and theoretical ratios, respectively, and $\delta R$ is the ratio uncertainty. The reduced $\chi^2$ quantifies the agreement between theory and observations: perfect agreement (within the adopted uncertainties) gives a value of 1.0; discrepancies beyond the stated uncertainties lead to $\chi^2 > 1$; and a reduced $\chi^2\le 1$ indicates that the adopted uncertainties have been overestimated.

\begin{figure}
  \scalebox{0.7}{\includegraphics{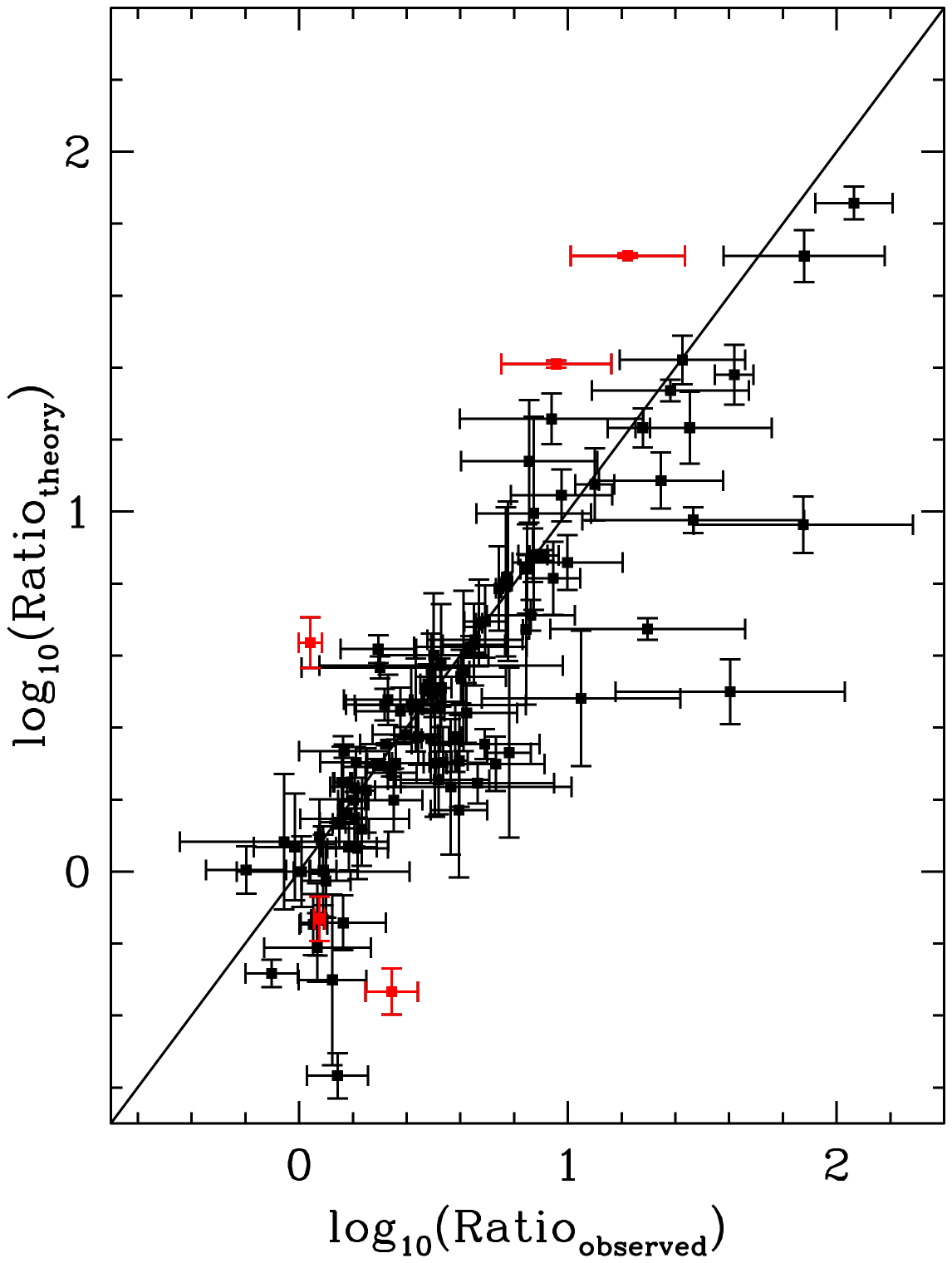}}
  \caption{Comparison of theoretical and observed intensity ratios for lines arising from the same upper level.}
  \label{fig:bratios}
\end{figure}

The first thing to note from Table~\ref{table:chis} is that the observational uncertainties by themselves are in general insufficient to account for the discrepancies between theoretical line ratios and observations. 
We see that the recent CIV3(DH11) \citep{deb11} calculation is marginally better than the previous computation by \citet{qui96}. The present NewTFDAc results look someone better than previous calculations. This seems contrary to the lifetimes comparisons of Table~\ref{table:lifetcomp}. This is because absolute A-values and radiative line widths are a lot more difficult to compute accurately. Thus, it is a reasonable practice to measure lifetimes experimentally and use them to correct A-values, while keeping the theoretical branching ratios. 
The `7-config' calculation, which employs the smallest expansion considered here, does yield the best agreement with observed spectra. Our recommended values seem to be a significant improvement over the older theoretical data, and once the theoretical uncertainties are taken into account, reasonable agreement with observations is reached. It may be appreciated that the branching ratios uncertainties are smaller than the uncertainties in the radiative rates, and if the latter are adopted $\chi^2=0.87$, thus indicating that the uncertainties are overestimated. The larger departures between theory and observations are associated to five line ratios: $\lambda 4458.5/\lambda 4515.5$ ($I(33{-}2)/I(33{-}3)$); $\lambda 4320.1/\lambda 4147.6$ ($I(39{-}8)/I(39{-}6)$); $\lambda 4373.0/\lambda 4147.6$ ($I(39{-}9)/I(39{-}6)$); $\lambda 16769/\lambda 15335$ ($I(11{-}7)/I(1{-}6)$); and $\lambda 4515.5/\lambda 4775.3$ ($I(33{-}3)/I(33{-}6)$). Not surprisingly, four of these problematic ratios involve level 6 ($a\ ^4F_{9/2}$) which is very difficult to represent theoretically. If these five ratios are removed, the reduced $\chi^2=1.07$. It is unclear if the remaining line-ratio discrepancies arise from observations or theory; nonetheless, as shown in Figure~\ref{fig:bratios}, the overall agreement between theory and observations is satisfactory.

\section{Electron Impact Collision Strengths}
\label{omega}

For the different \ion{Fe}{2} target expansions described above, collision strengths have been computed by two different methods, specifically R-matrix+ICFT
(RM+ICFT) and DARC. 
In the case of the small 7-configuration expansion it was also possible to do a Breit-Pauli calculation (BPRM), which compared very well with the 
RM+ICFT results. 
 We have also carried out calculations with different close-coupling expansions in the {\sc stg2} $R$-matrix code, with and without energy-corrected excitation thresholds at the Hamiltonian matrix diagonalization in {\sc stg3}. This allows us to estimate the sensitivity of the collisional data to target representation as well as to the 
details of the scattering calculation. Our runs include 20 continuum orbitals for each angular momentum in the close-coupling expansion, and partial waves up to $L=10$ in $LS$-coupling and $J=14.5$ in $JJ$-coupling. Collision strengths are sampled at 10\,000 equally spaced energy points up to the highest excitation threshold, with a much coarser mesh beyond it ($3\times$threshold).

For our DARC calculations
the target orbitals and energy levels 
were generated using the Dirac-Hartree-Fock atomic
structure program GRASP0 \citep{dya89, Parpia1996}. We employed only 
a minimal set of 6 configurations, including $3d^7,3p^43d^9,3p^63d^9,3d^64s,3d^64p$ and $3d^54s^2$ for a total of 329 
levels. The first 20 levels were shifted to the NIST values.  
The scattering calculations were performed using
our set of parallel Dirac $R$-matrix programs \citep{Ballance2006}, which merges 
of modified versions of the original codes developed by 
\cite{nor04} with our suite of
parallel Breit-Pauli $R$-matrix programs \citep{Ballance2004,
Mitnik2003}. 
The size of the $R$-matrix 'box' was 13.3 au and we employed 12 basis orbitals
for each continuum-electron angular momentum.  This was more than sufficient to
span electron energies up to 6.0 Ryd (81 eV). As the model considers only 
low temperature modeling, we only included partial waves up to J=10 with 
a top-up procedure to account for higher partial wave contributions \cite{burgess}, which 
were minimal. The cross sections spanned the energy range from the ground state 
to just over 1 Ryd, with an energy resolution of 10,000 points.

We computed thermally averaged effective collision strengths 
\begin{equation}
\Upsilon_{ij}=\int_0^\infty \Omega_{ij}(E_j)\exp({-E_j/k T_e})d(E_j/k T_e),
\end{equation}
where $\Omega_{ij}(E_j)$ and $E_j$ are the collision strength and incident electron energy relative to the $j$th level, respectively, $T_e$ is the electron temperature, and $k$ the Boltzmann constant.

Effective collision strengths at $10^4$~K from different calculations are compared in Table~\ref{table:upsil104}. Only a sub-set of  transitions from 
the ground level are shown, but the complete table for transitions among all the 52 levels is available electronically. Also, we do not show the results of all different calculations, but only the most representative ones.
The mean and standard deviation of the collision strength for each transition are also listed.
The most relevant comparisons involve excitations from levels of the ground and first excited multiplets since they dominate the whole spectrum, while most other collisional transitions among excited levels have very little impact on the resultant spectrum. 
Also, the comparison of effective collision strengths at $10^4$~K is perhaps the best way to make sense of the collision strength variations since this is the temperature where \ion{Fe}{2} is most frequently found. In comparisons at higher temperatures, the differences tend to be smaller because the contribution of near-threshold resonances to the collision strengths is less conspicuous. At lower temperatures the differences are enhanced because the near-threshold resonance structures dominate, but their exact position and interference patterns are uncertain. In fact, it is very difficult to provide effective collision strengths at temperatures below $\sim 5000$~K with reasonable confidence.

The first four columns of collision strengths in Table~\ref{table:upsil104} show the results of RM+ICFT calculations using a target with the same configurations as in \cite{qui96}. 
We refer to these calculations as 'Q+RM'. For this target, we look at the results when the theoretical target energies are used through the calculation (Q+RM-ns), the level energies are shifted in the H matrix to experimental values (Q-RM-shift), the
radius of the R-box is limited to 8 a.u. (Q+RM-RA=8), and the radius of the R-box is extended to 14.5 a.u. (Q+RM-RA=14.5). 
The next set of results corresponds to an RM+ICFT calculation using the '7-config' target described in Section 2. For this calculation we chose not to correct the target energies in {\sc stg3}.
The sixth set of effective collision strengths shown is from our best DARC calculation.

\begin{figure}
  \scalebox{0.6}{\includegraphics{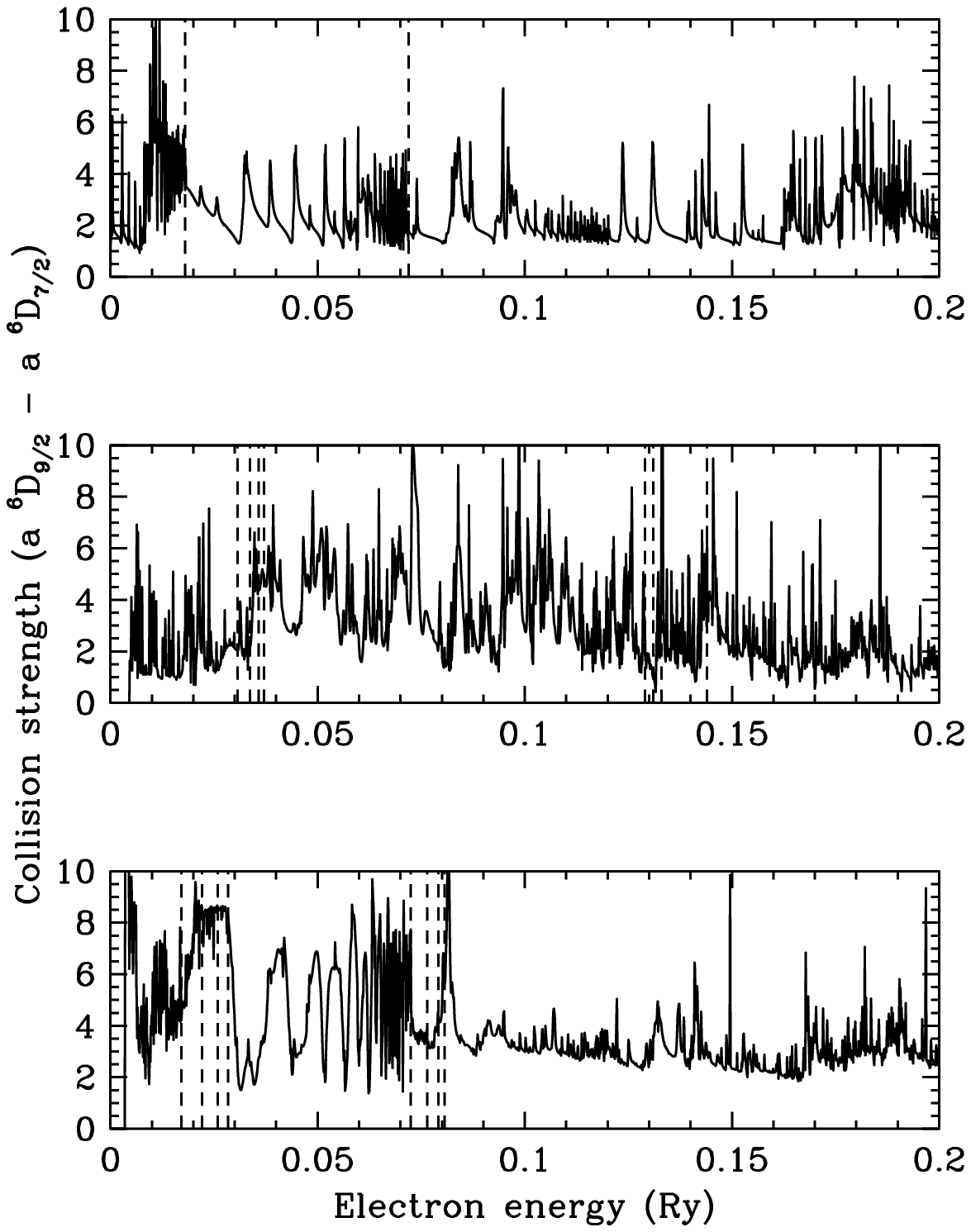}}
  \caption{Collision strengths for the 3d$^6$4s $^6$D$_{9/2}$ (ground level) to the $^6$D$_{7/2}$ (first excited level). The top panel shows the collision 
strength from a LS calculation with algebraic fine structure splitting using the `NewTDFAc' target, which gives accurate threshold energies. 
The  second and third panels show the collision strengths from our DARC calculations, keeping the theoretical energies and shifting the  thresholds to laboratory energies, respectively. The positions of the 3d$^7$~$^4$F and 3d$^6$4s~$^4$D thresholds are indicated in each panel by vertical dashed lines}
  \label{fig:omegafig}
\end{figure}

In addition to the calculations described above we did other calculations employing larger target expansions, but the results were neither significantly different from those published before nor yielded predicted spectra that compare favorably 
with experiment (see the next section).
In comparing the results of various calculations we find significant systematic differences in the collision strengths for transitions among levels of the $3d\,^64s\ ^6D$ ground multiplet.
For instance, for the excitation from the ground level to the first excited level, most of our current calculations give $\Upsilon(10^4 K)$ of about 2, while the {\sc darc} calculation and previously published works give $\Upsilon(10^4 K)$ of about 5.
The differences are found not to depend on the 
configuration expansion. Instead, the effective collision strengths are greatly enhanced when the excitation threshold are shifted to experimental values. This is illustrated in Figure~\ref{fig:omegafig}, where we plot the DARC collision strengths with and without energy corrections.
It can be seen that by shifting the target energies the Rydberg series of resonances seem to be pushed against the excitation threshold, then blended 
into near-threshold packs of resonances. Similar packs of resonances are seen in the the calculations of \cite{zhang95} and \cite{ram07}. 
In Figure~\ref{fig:omegafig} we also show the collision strengths using the ``NewTDFAc", which yields very accurate energies for the lowest \ion{Fe}{2} terms without corrections. It is found that the ``NewTDFAc" target does not give the large packs of resonances found in other calculations when the target energies are shifted to experimental values. While shifting threshold energies to experimental values is a technique generally used to improve the cross sections, such shifts are expected to be small. However, most target representations of \ion{Fe}{2} yield very poor energies for the first few excited terms of the system, thus large corrections are needed. 
While it is unclear whether the `NewTDFAc' target yields accurate collision  strength for these transition, it does show that previously published collision strengths for the ground multiplet are much more uncertain than previously thought.

By comparing the ÔQ+RM-nsÕ and ÔQ+RM-shiftÕ results that shifting the energies to experimental values has a very large effect for several transitions to highly excited levels. There are two reasons for this. Firstly, shifting the target energies also shifts the Rydberg series of resonances, particularly those resonances that lie near excitation threshold. Secondly, excited metastable levels with equal $j$ quantum numbers can be strongly mixed, and such mixing is greatly affected their energy separation. Moreover, when the energy shifts result in changing the relative order of such levels this can lead to large errors in the couplings that yield the resonances on the cross sections for such levels.

By comparing columns two, four, and five one finds that changing the size of the R-matrix box has a smaller effect on the cross sections than shifting the target energies, yet the effect is still sizable for a few transitions. These effects should also be considered in assessing the uncertainty in the cross sections.

It should be stressed that all calculations generally agree on background cross sections, and the differences are mostly due to the resonance structures. Furthermore, we take into account all results, and their statistical dispersion is assumed an accuracy indicator.

Regarding the overall error distribution, it is found that, for excitations from the lower nine levels of the ion that dominate the entire spectrum, uncertainties in the range of 10--20\% are the most frequent. This is closely followed by those in the 20--30\% range. Moreover, most of the effective collision strengths have uncertainties less than 50\%, with a tendency for the smaller collision strengths to be poorer than the weaker ones. If the complete collisional inventory is considered, the effective-collision-strength discrepancies vary widely, reaching factors of 2 or more in some cases. 
Despite these bulk uncertainties in the collision strengths the strong transitions that dominate the spectrum under typical nebular conditions seem to be reasonably well known,  as will be discussed in the next section.

It should be noted that because the largest atomic data uncertainties are in the collision strengths it is expected that modeled spectra will be more accurate as the electron density increases, while larger error are expected in models of very diluted plasmas (see Bautista et al. 2013).

The current data set establishes a firm foundation from which specific transitions of interest can be studied individually in order to reduce their current uncertainties. However, it should not be assumed {\em de facto} that the larger expansions yield the more accurate results unless they are proven to be fully converged.

%\begin{figure}\resizebox{\hsize}{!}
%{\includegraphics{intensityfig.eps}}
%\caption{Comparison between predicted line emissivities from our present
%best data and data from \citep{zhang95} (effective collision strengths) and
%\citep{naharpradhan} (A-values) and measured intensities of
%lines from the "shock component" in the Orion nebulae \cite{mesadelgado}.}
%\label{fig:intensity}
%\end{figure}

\section{Comparison with observed spectra}
\label{spectra}

In order to assess the collisional data quality, we have constructed collisional excitation models with the recommended $A$-values from Section~\ref{rad} and the effective collision strength data sets introduced in Section~\ref{omega}. The predicted emission spectra for each of these models, for several electron temperatures and densities, are compared with the spectra of the Herbig--Haro object HH~202 \citep{mes09} and the pre-main-sequence star SEO-H$\alpha$~574 \citep{gia13}. 
All [\ion{Fe}{2}] lines measured in these objects are inlcuded in the comparisons, i.e. 78 and 55 respectively. We make no attempt to compare
with the [\ion{Fe}{2}] spectrum of the Weiglet blobs of $\eta$ Carinae because the strong UV \ion{Fe}{2} emission from this object indicates that there is significant fluorescent excitation. Atomic rate error propagation in predicted line emissivities has been examined by \citet{bau13}; for the present comparisons, we adopt the statistical scatter of the $A$-values and effective collision strengths at $10^4$~K as estimates of the atomic data error  for each transition.
Instead of comparing absolute line intensities which depend on ion abundances, we normalize the observed lines to the sum of all line fluxes and the theoretical emissivities to the sum of the
line emissivities of all lines in the observed spectrum, and we vary the electron temperature and density in each model to get the best possible accord with observations. The theory--experiment correlation is expressed in terms of the reduced $\chi^2$.

Table~\ref{table:linecomps} shows the model physical conditions and reduced $\chi^2$ values that best match the observed HH~202 spectrum, where each model implements a specific set of effective collision strengths. As gauged by the reduced $\chi^2$ index, 
it may be seen that the model with the effective collision strengths compiled by \citet{baup98} leads to a fairly poor correlation with the observed spectrum.

Model RH07, that employees the effective collision strengths of \citet{ram07} and \citet{ram09} fairs even worse than previous models when compared with optical spectra. 

The collision strengths computed with the fully relativistic DARC code yield the worst agreement with observation. This is to be expected, since it was not possible to construct a self-consistent 4d orbital for this target, that accounted for
important relaxation effects of the spectroscopic 3d orbital.

The RM+ICFT calculations shown in Table~9 use the same configurations as the `New-HFR' for the radiative calculations. This target expansion is used for two calculations with 63 and 114 levels in the close coupling expansion. The
two results yield line intensities that agree significantly better with observations than previous calculations.

\begin{figure}
  \scalebox{0.6}{\includegraphics[angle=-90]{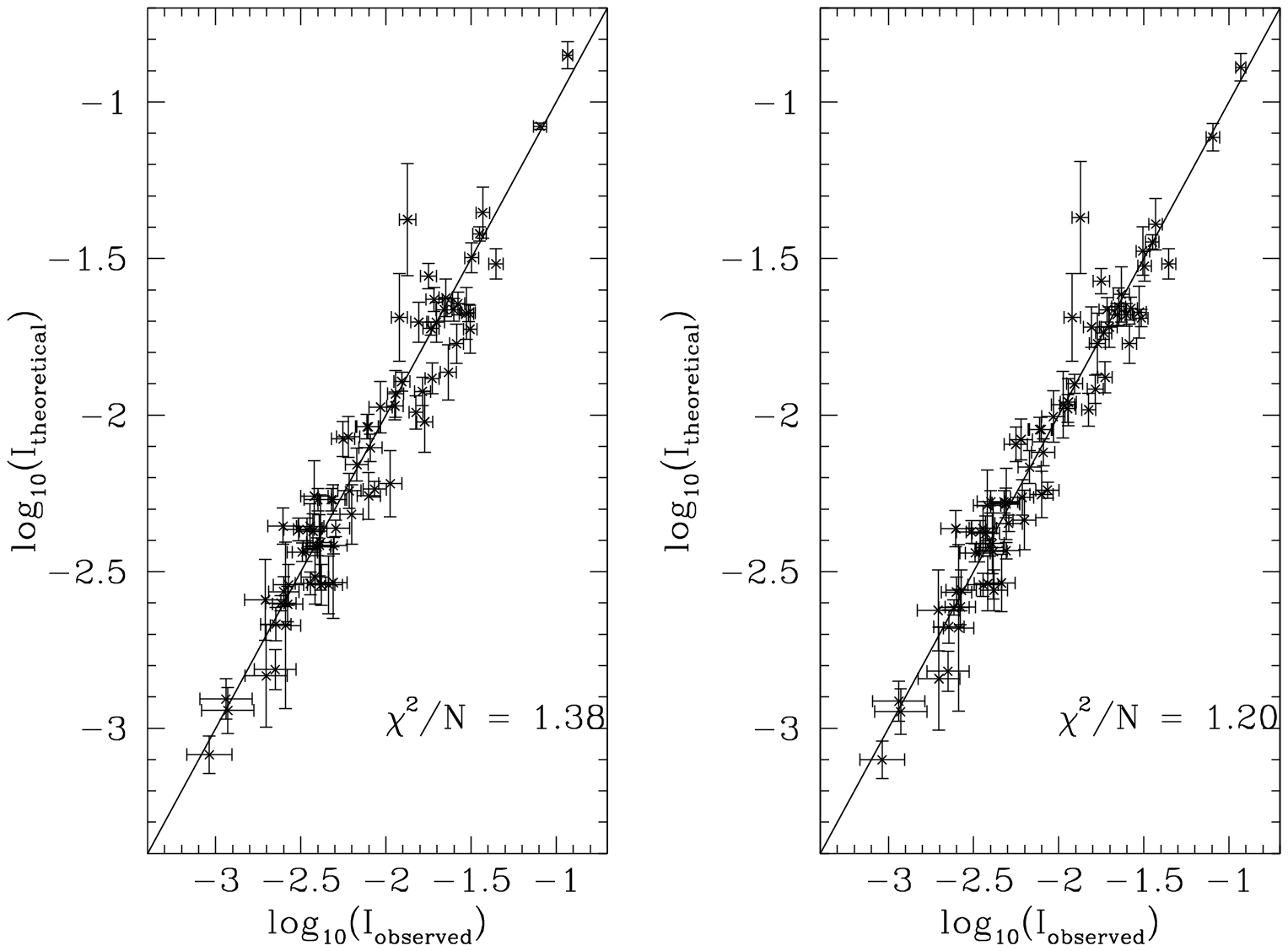}}
  \caption{Comparison between observed and predicted normalized line strengths for two different objects.}
  \label{fig:comparespec}
\end{figure}

The observed spectrum is best reproduced with the collision strengths of the `7-config.'. With respect to the HH~202 spectrum, this collisional data set yields a reduced $\chi^2$ of 1.01 for $T_e=11\,500$~K and $n_e =6.56\times 10^4$~cm$^{-3}$; i.e. a nearly perfect match. It is interesting that the \ion{Fe}{2} spectrum points to an electron density that is roughly four times the density diagnosed by \cite{mes09} with higher ionization species. This is consistent with an observed shocked region where the \ion{Fe}{2} emission arises from the shock front itself. Figure~\ref{fig:comparespec} depicts a comparison between predicted and observed spectra.

As good as the agreement with the optical spectrum of HH~202 is, it does not constrain the collision strengths among the lowest 16 levels of \ion{Fe}{2},
which yield infrared lines. Thus, we look at the spectrum of SEO-H$\alpha$~574, which combines optical and near-IR lines. 
The
 between 1 to 2 m, includes 14 lines from
de-excitations from the 3d$^6$4s a~$^4$D multiplet to the 3d$^6$4s a~$^6$D and 3d$^7$ a~$^4$F
terms. 
We find that the best agreement
with observation is found when combining the RH07 collision strengths for the lowest 16 levels and the `7-config' collision strengths for the higher levels.
This model yields a reduced $\chi^2$ of 1.32 fort $T_e=10\,000$~K and $n_e =4.4\times 10^4$~cm$^{-3}$. 
It is apparently clear that the RH07 model yields more
accurate collision strenghts than our preferred model for transitiona among
the first three terms of Fe II, and particularly between the a~$^6$D and a~$^4$D
terms of the 3d$^6$4s configuration. This may be due to the way the target
wavefunctions were optimized in RH07. Yet, this same accuracy does not seem
to be maintained through higher levels.

The agreement with SEO-H$\alpha$~574 is worse than for HH~202 due in part, we believe, to the scantier number of spectral lines (only 53 lines) and to 
underestimated observational uncertainties. In published spectra, it is claimed that nearly a third of the lines are accurate to better than 5\% and almost 
two-thirds to better than 10\%. Given the lower resolution of the SEO-H$\alpha$~574 spectrum, relative to that of HH~202, the trial uncertainties in the 
measured fluxes could be roughly twice as large as those claimed, in which case the agreement with the theoretical prediction would be comparable to that found for HH~202.

\subsection{Emission spectra: IR and optical}

We examine all lines longwarth of 3~$\mu$m since different instruments cover different segments of the mid-IR and IR regions,
the most important [\ion{Fe}{2}] lines in increasing wavelength order being at: 5.34~$\mu$m ($3d\,^64s\ ^6D_{9/2} - 3d\,^7\ ^4F_{9/2}$; 1~--~6); 17.93~$\mu$m ($3d\,^7\ ^4F_{9/2} - 3d\,^7\ ^4F_{7/2}$; 6~--~7); 22.92~$\mu$m ($3d\,^64s\ ^4D_{7/2}- 3d\,^64s\ ^4D_{5/2}$; 10~--~11);
24.51~$\mu$m ($3d\,^7\ ^4F_{7/2} - 3d\,^7\ ^4F_{5/2}$; 7~--~8); 25.98~$\mu$m ($3d\,^64s\ ^6D_{9/2} - 3d\,^64s\ ^6D_{7/2}$; 1~--~2);
35.34~$\mu$m ($3d\,^64s\ ^6D_{7/2} - 3d\,^64s\ ^6D_{5/2}$; 2~--~3); and 35.77~$\mu$m ($3d\,^64s\ ^6D_{5/2} - 3d\,^64s\ ^6D_{3/2}$; 8~--~9).

\begin{figure}
  \scalebox{0.6}{\includegraphics[angle=-90]{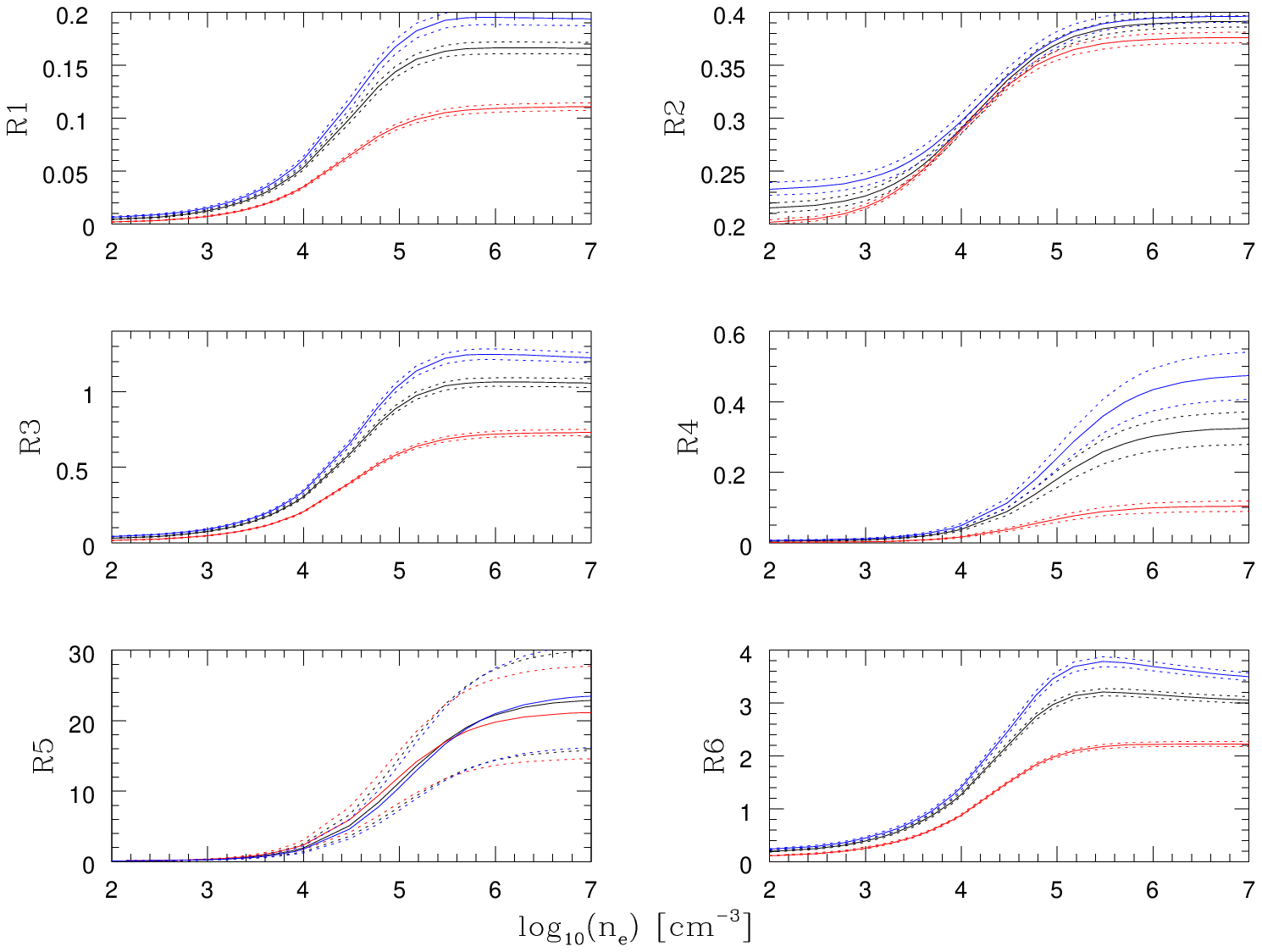}}
  \caption{IR density diagnostic line ratios. Line-ratio errors resulting from the atomic data are indicated by the dotted lines. The ratios are calculated at temperatures of 5000~K (lowest, red lines), 10\,000~K (middle, black lines), and 15\,000~K (upper, blue lines). The different ratios are:
  $R1= j(35.77\,\mu{\rm m})/j(25.98\,\mu{\rm m})$;
  $R2= j(35.34\,\mu{\rm m})/j(25.98\,\mu{\rm m})$;
  $R3= j(24.52\,\mu{\rm m})/j(25.98\,\mu{\rm m})$;
  $R4= j(22.89\,\mu{\rm m})/j(25.98\,\mu{\rm m})$;
  $R5= j(17.93\,\mu{\rm m})/j(5.34\, \mu{\rm m})$; and
  $R6= j(17.93\,\mu{\rm m})/j(25.98\,\mu{\rm m})$.}
  \label{fig:ratiosIR}
\end{figure}

Figure~\ref{fig:ratiosIR} shows emissivity line ratios among these lines as a function of electron density for temperatures between $5000{-}15\,000$~K. These ratios can be used as diagnostics of electron density in the range of $10^3{-}10^5$~cm$^{-3}$. Ratios that involve the 5.34~$\mu$m line carry a significant error, particularly toward the high densities due to the 30\% uncertainty in the lifetime of the $3d\,^7\ ^4F_{9/2}$ level. All other line ratios are well constrained and should provide reliable diagnostics. The $j$(35.34~$\mu$m)/$j$(25.98~$\mu$m) ratio is particularly relevant because it is essentially invariant with temperature; therefore, this ratio in combination with any of the other IR ratios would constrain density and temperature in the [\ion{Fe}{2}] emitting region.

There is a good number of [\ion{Fe}{2}] lines in the near-IR region ($1{-}3$~$\mu$m) that originate from radiative transitions involving levels within the $3d\,^64s\ ^4D$ multiplet. Having more than one line from the same upper level is useful as they can be used as a dust-extinction diagnostic. The stronger lines in this part of the spectrum are at 1.257~$\mu$m ($3d\,^64s\ ^6D_{9/2} - 3d\,^64s\ ^4D_{7/2}$; 1~--~10) and 1.644~$\mu$m ($3d\,^7\ ^4F_{9/2} - 3d\,^64s\ ^4D_{7/2}$; 6~--~10). The intrinsic $j$(1.644~$\mu$m)/$j$(1.257~$\mu$m) ratio is estimated to be 0.80 with an accuracy of $\sim 20\%$, but extinction due to dust would increase this ratio in observed spectra.

\begin{figure}
  \scalebox{0.6}{\includegraphics[angle=-90]{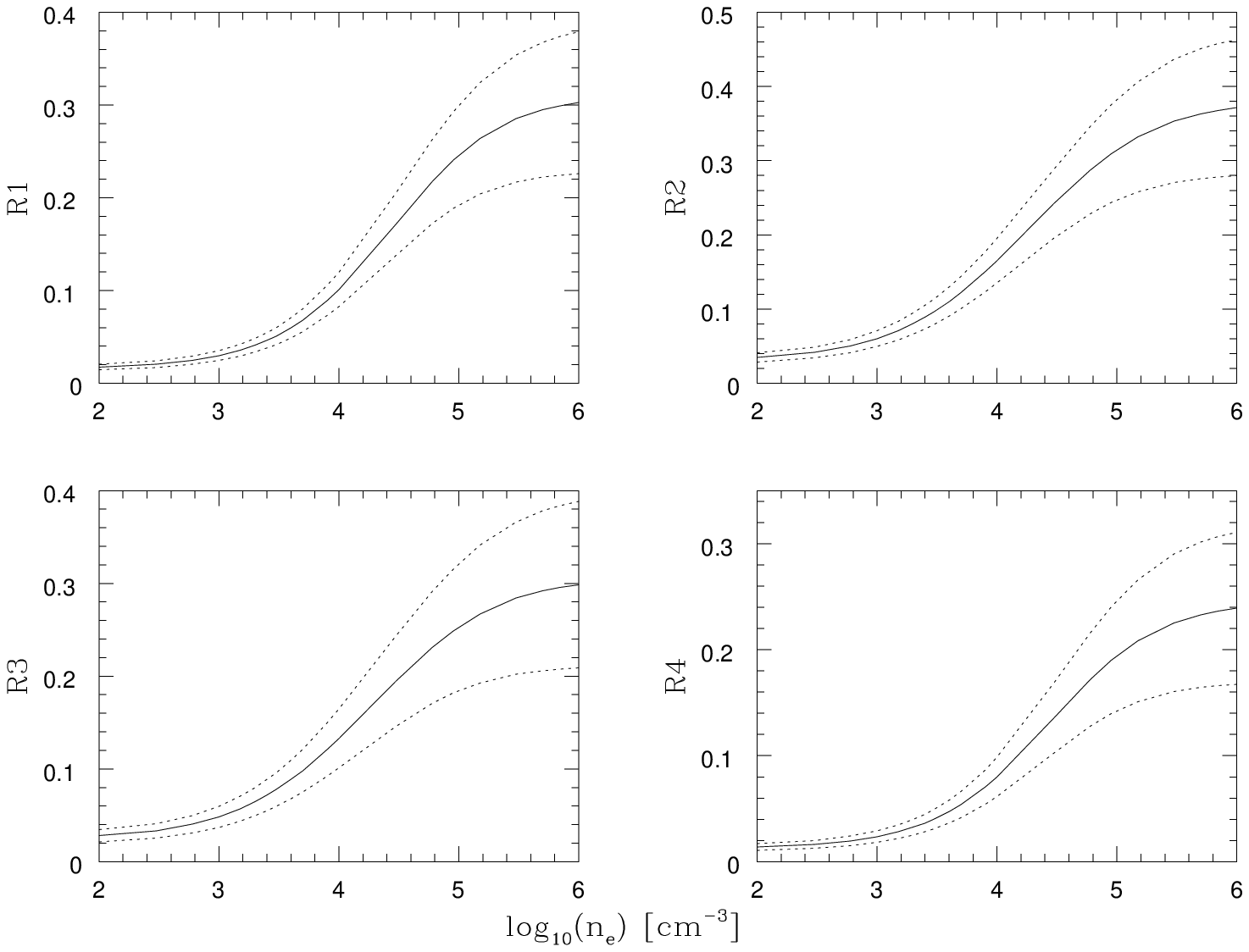}}
  \caption{Near-IR density diagnostic line ratios at 10\,000~K. Line-ratio errors resulting from the atomic data are indicated by the dotted lines. The different ratios are:
  $R1= j(1.600\, \mu{\rm m})/j(1.644\, \mu{\rm m})$;
  $R2= j(1.534\, \mu{\rm m})/j(1.644\, \mu{\rm m})$;
  $R3= j(1.294\, \mu{\rm m})/j(1.257\, \mu{\rm m})$; and
  $R4= j(1.279\, \mu{\rm m})/j(1.257\, \mu{\rm m})$.}
  \label{fig:ratiosnIR}
\end{figure}

The near-IR line ratios are insensitive to temperature variations at $\sim 10^4$~K, but they can be used as density diagnostics in the range $10^3{-}10^{4.5}$~cm$^{-3}$. Such line ratios reach the high-density limits beyond $10^5$~cm$^{-3}$, but near those limits, the ratios are of little use due to the current uncertainties in the radiative branching ratios as illustrated in Figure~\ref{fig:ratiosnIR}. Moreover, observations of dense nebulae in the near-IR could be very useful to constrain the high-density limits of various diagnostics and, hence, the transition branching ratios.

\begin{figure}
  \scalebox{0.6}{\includegraphics[angle=-90]{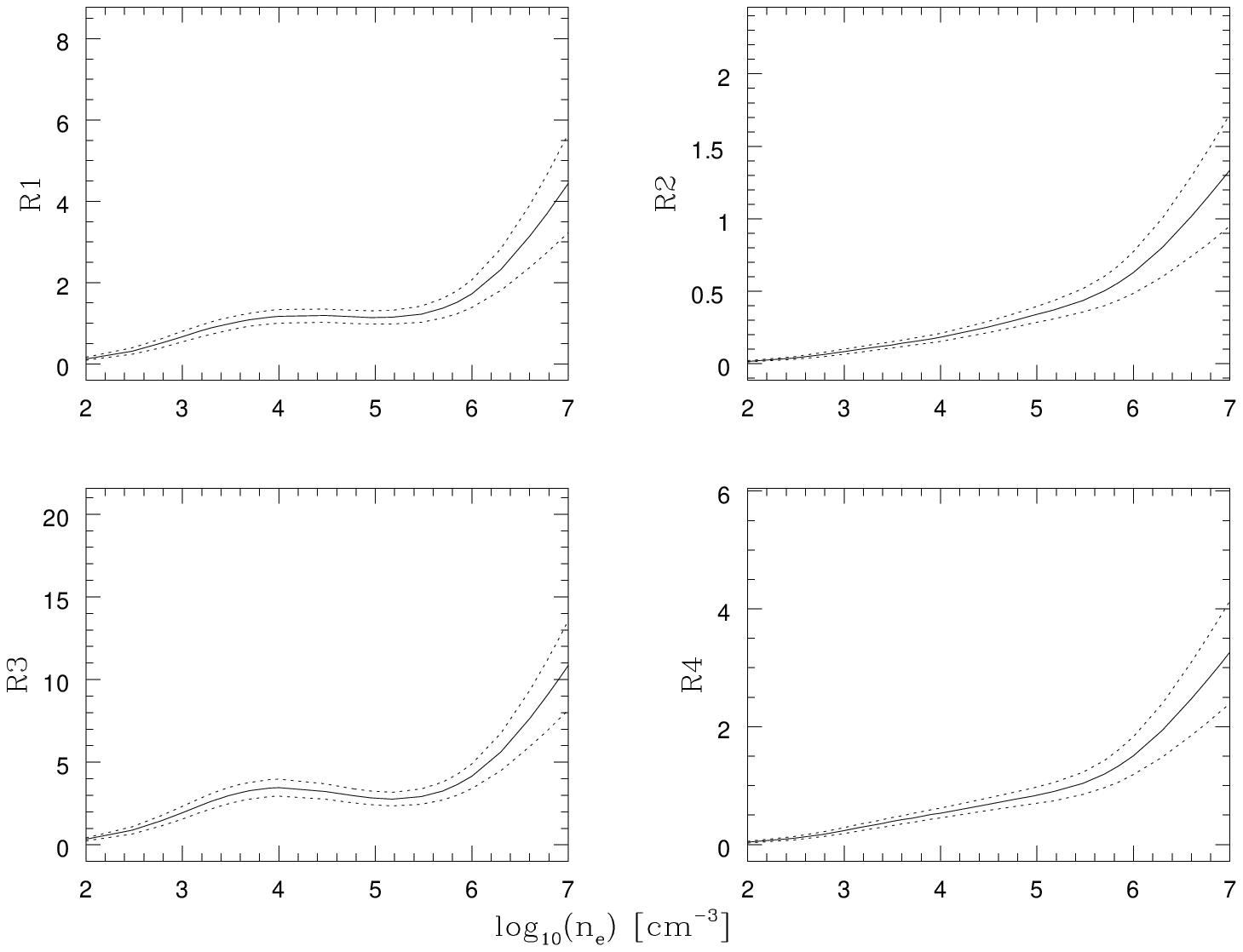}}
  \caption{Red density diagnostic line ratios at 10\,000~K. Line-ratio errors resulting from the atomic data are indicated by the dotted lines. The different ratios are:
  $R1= j(7155.2$~\AA)$/j(8616.8$~\AA);
  $R2= j(7452.6$~\AA)$/j(8616.8$~\AA);
  $R3= j(7155.2$~\AA)$/j(8891.8$~\AA); and
  $R4= j(7452.6$~\AA)$/j(8891.8$~\AA).}
  \label{fig:ratiosred}
\end{figure}

The stronger lines in the red part of the spectrum ($7400{-}10\,000$~\AA) result from de-excitations of the $3d\,^7\ ^4P$ and $^2G$ levels. There are four particularly strong lines whose strengths are accurately determined both theoretically and observationally, and are in good agreement in the spectrum of HH~202: 8616.8~\AA\ ($3d\,^7\ ^4F_{9/2} - 3d\,^7\ ^4P_{5/2}$; 6~--~14); 8891.8~\AA\ ($3d\,^7\ ^4F_{7/2} - 3d\,^7\ ^4P_{3/2}$; 7~--~15); 7155.2~\AA\ ($3d\,^7\ ^4F_{9/2} - 3d\,^7\ ^2G_{9/2}$; 6~--~17); and 7452.6~\AA\ ($3d\,^7\ ^4F_{7/2} - 3d\,^7\ ^2G_{7/2}$; 7~--~17). From these lines four diagnostic ratios can be determined that cover the range $10^2{-}10^7$~cm$^{-3}$; for densities higher than $10^7$~cm$^{-3}$, the ratios are currently too uncertain for any diagnostic purposes. Useful ratios are shown in Figure~\ref{fig:ratiosred}.

\begin{figure}
  \scalebox{0.6}{\includegraphics[angle=-90]{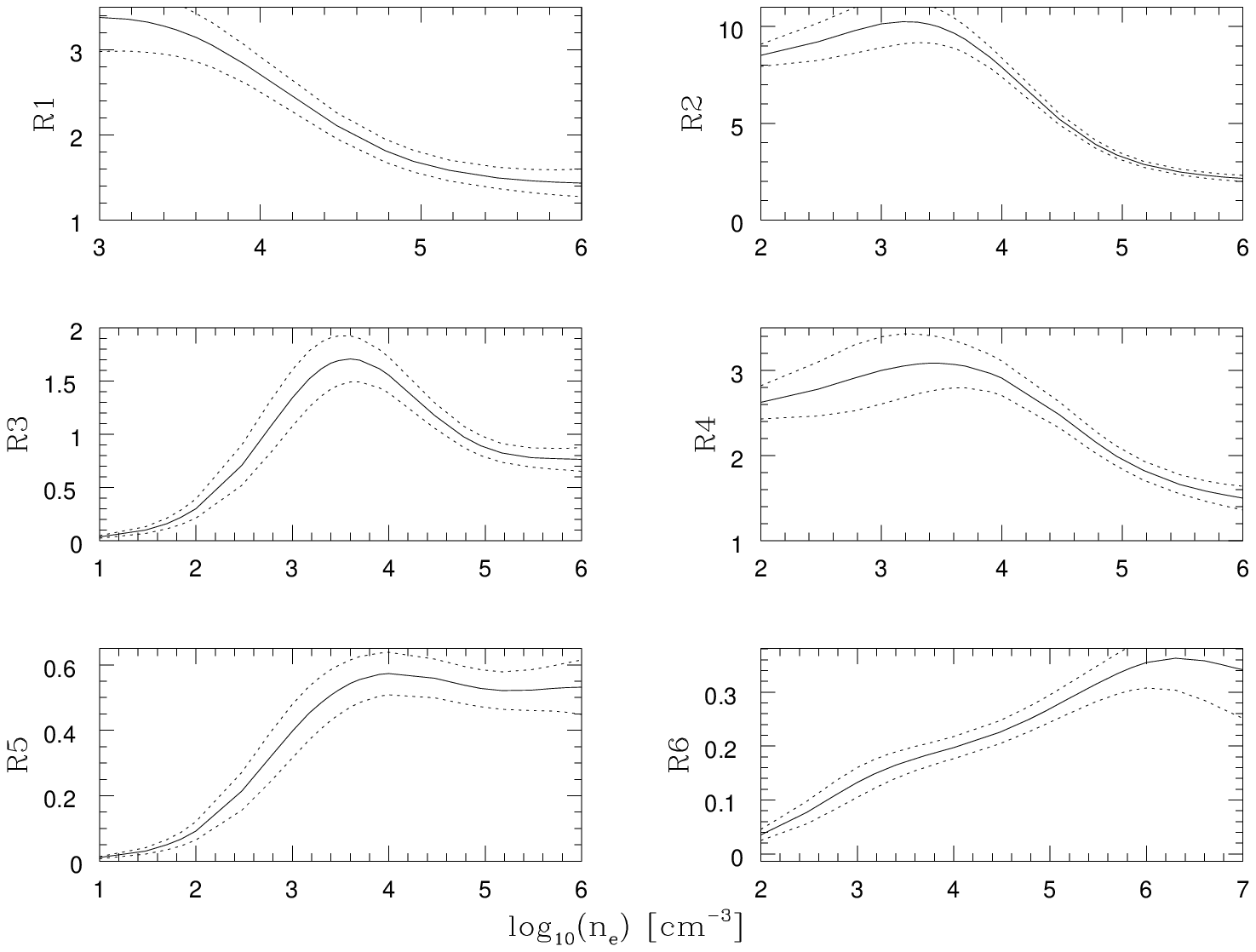}}
  \caption{Optical density diagnostic line ratios at 10\,000~K. Line-ratio errors resulting from the atomic data are indicated by the dotted lines. The different ratios are:
  $R1= j(5261.6$~\AA)$/j(5333.6$~\AA);
  $R2= j(5158.0$~\AA)$/j(5333.6$~\AA);
  $R3= j(5527.4$~\AA)$/j(5333.6$~\AA);
  $R4= j(5158.0$~\AA)$/j(5261.6$~\AA);
  $R5= j(5527.4$~\AA)$/j(5261.6$~\AA); and
  $R6= j(5158.0$~\AA)$/j(5527.4$~\AA).}
  \label{fig:ratiosOP}
\end{figure}

The visible [\ion{Fe}{2}] spectrum ($4000\le\lambda\le 7400$~\AA) is very rich, exhibiting lines from excited levels
$\sim2.5{-}3.8$~eV above the ground level. We have selected the seven strongest and most reliable visible lines for diagnostic ratios: 5527.4~\AA\ ($3d\,^7\ ^4F_{7/2} - 3d\,^7\ ^2D_{5/2}$; 7~--~27); 5158.8~\AA\ ($3d\,^7\ ^4F_{7/2}-3d\,^64s\ ^4H_{13/2}$; 7~--~26);
5261.6~\AA\ ($3d\,^7\ ^4F_{7/2} - 3d\,^64s\ ^4H_{11/2}$; 7~--~29); 5111.6~\AA\ ($3d\,^7\ ^4F_{9/2}-3d\,^64s\ ^4H_{11/2}$; 6~--~29);
5333.6~\AA\ ($3d\,^7\ ^4F_{5/2}- 3d\,^64s\ ^4H_{9/2}$; 8~--~30); 5220.0~\AA\ ($3d\,^7\ ^4F_{7/2}-3d\,^64s\ ^4H_{9/2}$; 7~--~30); and 4745.5~\AA\ ($3d\,^7\ ^4F_{9/2}- 3d\,^74s\ ^4F_{5/2}$; 6~--~34). The most useful line-ratio diagnostics are shown in Figure~\ref{fig:ratiosOP}, which can be used in a wide density range ($10{-}10^6$~cm$^{-3}$) and are mostly insensitive to temperature variations around $\sim 10\,000$~K.

\begin{figure}
  \scalebox{0.6}{\includegraphics[angle=-90]{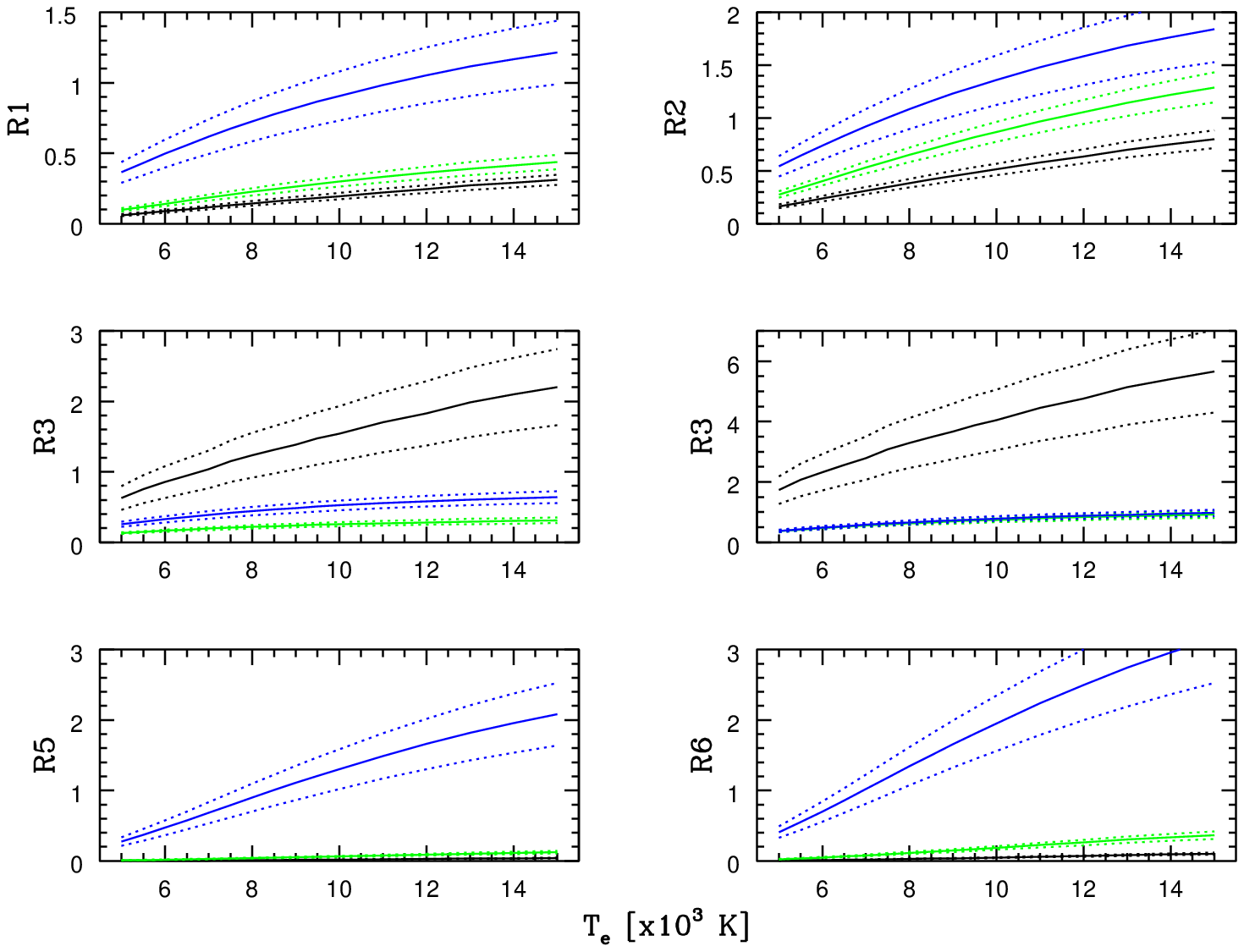}}
  \caption{Temperature diagnostic line ratios combining optical and red or near-IR lines. The ratios are plotted for densities of 100~cm$^{-3}$ (black lines), $10^4$~cm$^{-3}$ (green lines), and $10^6$~cm$^{-3}$ (blue lines). Line-ratio errors resulting from the atomic data are indicated by the dotted lines. The different ratios are:
  $R1= j(5261.6$~\AA)$/j(8616.0$~\AA);
  $R2= j(5158.0$~\AA)$/j(8616.0$~\AA);
  $R3= j(5261.6$~\AA)$/j(7155.2$~\AA);
  $R4= j(5158.0$~\AA)$/j(7155.2$~\AA);
  $R5= j(5261.6$~\AA)$/j(1.2567$~$\mu$m); and
  $R6= j(5158.0$~\AA)$/j(1.2567$~$\mu$m)}
  \label{fig:ratiosROP}
\end{figure}

The best temperature diagnostics are obtained from line ratios of optical, red, or near-IR lines. This is illustrated in Figure~\ref{fig:ratiosROP} using the stronger lines in each part of the spectrum. These ratios can be very useful in the analysis of spectra from instruments such as the X-Shooter spectrograph that covers visible and near-IR ranges simultaneously.

\subsection{Absorption spectra: UV transitions}

\begin{figure}
  \scalebox{0.6}{\includegraphics[angle=-90]{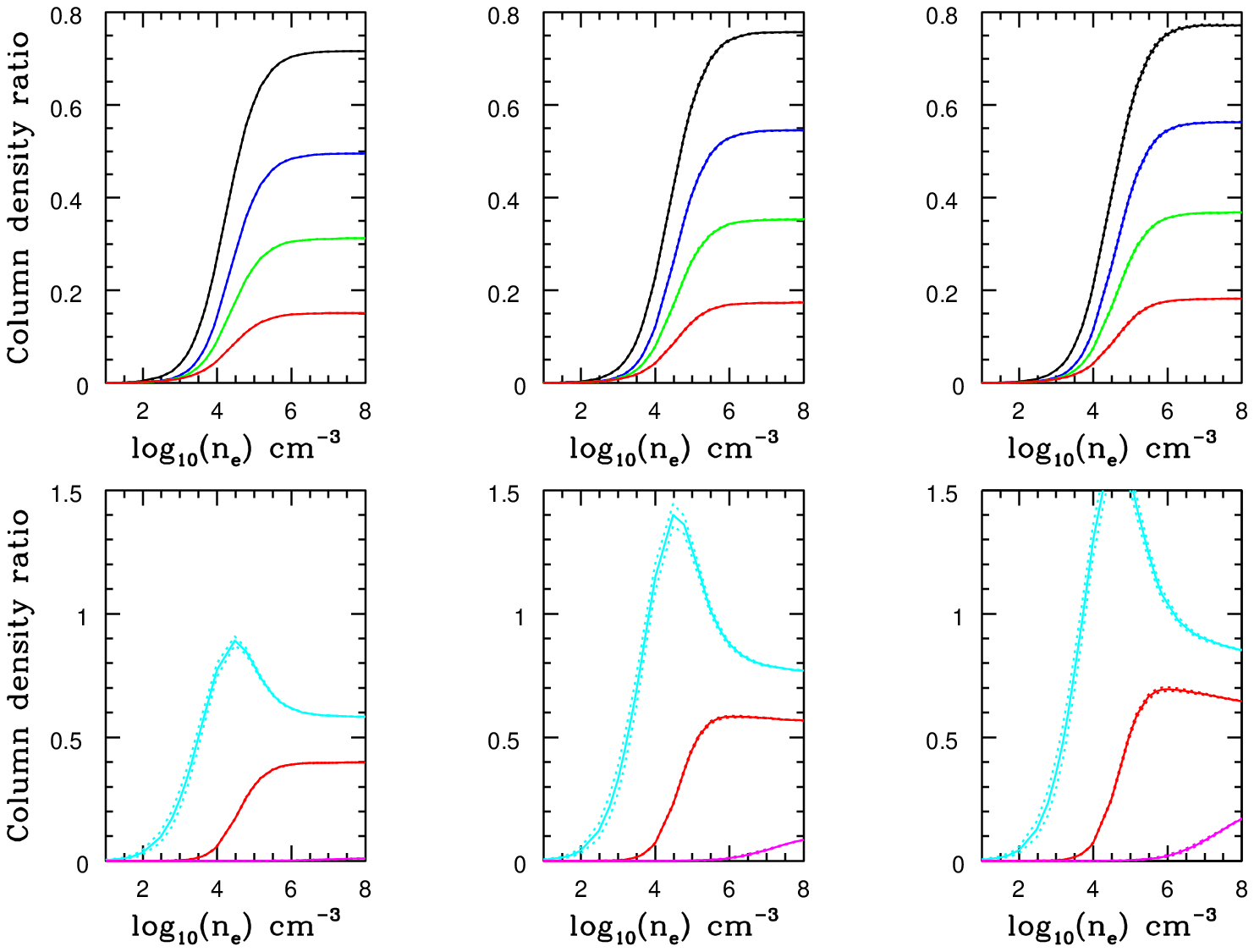}}
  \caption{Excited level column density (relative to that of the ground level) as a function of electron density. Results for temperatures of 5000~K, 10\,000~K, and 15\,000~K are listed from left to right. The top panels show the results for the four excited levels of the $^6D$ multiplet, starting with the level with $E=384.8$~cm$^{-1}$ (top curve) and ending with that with $E=977.0$~cm$^{-1}$ (lowest curve). In the lower panels we show the relative column densities of levels with $E=1872.6$~cm$^{-1}$ (top curve), $E=2430.1$~cm$^{-1}$ (middle curve), and $E=7955.3$~cm$^{-1}$ (lowest curve).}
  \label{fig:columns}
\end{figure}

UV absorption spectra of bright sources, e.g. AGN and GRB, often exhibit \ion{Fe}{2} troughs. In some objects it is possible
to measure troughs from excited levels together with the resonant transitions, and this offers the possibility to use measured

column densities as density and temperature diagnostics \citep[see, for example,][and references therein]{dun10}. At typical temperatures of $\sim 10^4$~K, the most populated levels and those that have been observed in absorption arise from the $3d\,^64s\ ^6D$ multiplet (with energies 0.0, 384.8, 667.7, 862.6, and 977.0~cm$^{-1}$) and the high-multiplicity levels of the $3d\,^7\ ^4F$ and $3d\,^64s\ ^4D$ multiplets (at 1872.6, 2430.1, and 7955.3~cm$^{-1}$). Figure~\ref{fig:columns} shows column density ratios for these levels as a function of electron density for three different temperatures. It can be seen that these column density ratios are distinct density diagnostics for densities up to $\sim 10^6$~cm$^{-3}$.

It is noted that the current results resolve the discrepancy found by \cite{dun10} in the column density from the 1873~cm$^{-1}$ level (a~$^4$F$_{9/2}$) in the spectrum of the FeLowBAL quasar SDSS~J0318-0600. In that work, it was found that the collision strengths of \citet{baup98} reproduced the observed column density for this level for a density $\approx 0.5$~dex lower than the estimates from other \ion{Fe}{2} and \ion{Si}{2} levels. Some fluorescence effects were put forward as possible explanations for this effect; however, the current atomic data, on there own merits, yield column densities much more consistent with observations than previous models.

\section{Discussion and Conclusions}

We present a complete spectral model for the \ion{Fe}{2} ion comprising the lowest 52 metastable levels of the ion. It accounts for essentially all dipole forbidden lines of the ion in the optical and IR spectral regions, as well as the column densities of absorption troughs in the UV. This model is the result from extensive revisions and calculations of atomic parameters, namely dipole forbidden $A$-values and electron impact collision strengths. For these calculations we employ several different state-of-the-art theoretical methods, and compare the results with previously published data. These comparisons allow us to estimate the uncertainty in each atomic rate, and propagate them through spectral model predictions.

A general conclusion that can be derived from all the different calculations is that---for such a complex atomic system as [\ion{Fe}{2}] and the very large number of energy levels and transitions involved in the spectrum---not a single calculation can achieve convergence so as to provide all accurate atomic parameters at once. Moreover, very large atomic expansions tend to give worse overall results than small, well-controlled, and optimized expansions.

Furthermore, we employ a number of astronomical observations with exceptionally rich [\ion{Fe}{2}] spectra as benchmarks for the atomic data. We find very good agreement between observations and our recommended data within the estimated uncertainties derived from measurements and calculated values. Moreover, our spectral model is able to predict optical emission spectrum of over 100 lines of the HH~202 object in the Orion nebula in nearly perfect statistical agreement with observations.

Our atomic model is then used to explore the most important [\ion{Fe}{2}] line diagnostic ratios in the optical, near-IR, and IR regions. We also present useful column density diagnostic ratios in the UV.

The atomic data, with estimated uncertainties, from this work is available in Table 4 (radiative life times), Table 10 (in electronic form; branching ratios), Table 11 (in electronic form; ratios from lines from the same upper level in $\eta$~Carinea), Table 11 (in electronic form, effective collision strengths), and Table 12 (in electronic form, estimated effective collision strength uncertainties).

\section{Acknowledgements}
We acknowledge support from the NASA Astronomy and Physics Research and Analysis Program (Award NNX09AB99G) and the Space Telescope Science Institute (Project GO-11745).
VF is currently a postdoctoral researcher of the Return Grant Program of the Belgian Scientific Policy (BELSPO). PQ is Research Director of the Belgian National Fund for Scientific Research F.R.S.-FNRS.

%%%% References

%%%%%%%%%%%%%%%%%%%%%%%%%%%%%%%%%%%%%%%%%%%%%%%%%%%%%%%%%%%%%%%%%%%%%%%%%%%%%%%

%%%%%%%%%%%%%%%%%%%%%%%%%%%%%%%%%%%%%%%%%%%%%%%%%%%%%%%%%%%%%%%%%%%%%%%%%%%%%%%

%%%%%%%%%%%%%%%%%%%%%%%%%%%%%%%%%%%%%%%%%%%%%%%%%%%%%%%%%%%%%%%%%%%%%%%%%%%%%%%

%%%%%%%%%%%%%%%%%%%%%%%%%%%%%%%%%%%%%%%%%%%%%%%%%%%%%%%%%%%%%%%%%%%%%%%%%%%%%%%

%%%%%%%%%%%%%%%%%%%%%%%%%%%%%%%%%%%%%%%%%%%%%%%%%%%%%%%%%%%%%%%%%%%%%%%%%%%%%%%

%%%%%%%%%%%%%%%%%%%%%%%%%%%%%%%%%%%%%%%%%%%%%%%%%%%%%%%%%%%%%%%%%%%%%%%%%%%%%%%

%%%%%%%%%%%%%%%%%%%%%%%%%%%%%%%%%%%%%%%%%%%%%%%%%%%%%%%%%%%%%%%%%%%%%%%%%%%%%%%

%%%%%%%%%%%%%%%%%%%%%%%%%%%%%%%%%%%%%%%%%%%%%%%%%%%%%%%%%%%%%%%%%%%%%%%%%%%%%%%

%%%%%%%%%%%%%%%%%%%%%%%%%%%%%%%%%%%%%%%%%%%%%%%%%%%%%%%%%%%%%%%%%%%%%%%%%%%%%%%

%%%%%%%%%%%%%%%%%%%%%%%%%%%%%%%%%%%%%%%%%%%%%%%%%%%%%%%%%%%%%%%%%%%%%%%%%%%%%%%

%%%%%%%%%%%%%%%%%%%%%%%%%%%%%%%%%%%%%%%%%%%%%%%%%%%%%%%%%%%%%%%%%%%%%%%%%%%%%%%

%%%%%%%%%%%%%%%%%%%%%%%%%%%%%%%%%%%%%%%%%%%%%%%%%%%%%%%%%%%%%%%%%%%%%%%%%%%%%%%

%%%%%%%%%%%%%%%%%%%%%%%%%%%%%%%%%%%%%%%%%%%%%%%%%%%%%%%%%%%%%%%%%%%%%%%%%%%%%%%

\newpage

\begin{deluxetable}{ll}
\tabletypesize{\scriptsize}
\tablecaption{{\sc autostructure} configuration expansions for \ion{Fe}{2}\label{table:AUTOconfigs}}
\tablewidth{0pt}
\tablehead{\colhead{Label} & \colhead{Configuration Expansion}}
\startdata
Spectroscopic &
$3s^23p^63d\,^64s$,
$3s^23p^63d\,^7$,
$3s^23p^63d\,^54s^2$ \\
\hline
\noalign{\smallskip}
BP extend TFDAc&
$3s^23p^63d\,^6\overline{4d}$,
$3s^23p^63d\,^64p$,
$3s^23p^63d\,^54s4p$,
$3s^23p^53d\,^74s$,
$3s^23p^53d\,^8$,
$3s^23p^43d\,^9$,\\
 & $3s3p^63d\,^74s$,
$3s3p^63d\,^74p$,
$3s3p^63d\,^8$,
$3p^63d\,^84s$,
$3p^63d\,^74s4p$,
$3p^63d\,^9$,
$3p^63d\,^74s^2$ \\
\hline
\noalign{\smallskip}
Q96+4d$^2$ &
$3s^23p^63d\,^6\overline{4d}$,
$3s3p^63d\,^74s$,
$3s3p^63d\,^8$,
$3s^23p^63d\,^54p^2$,
$3s^23p^63d\,^5\overline{4d}^2$,
$3s^23p^63d\,^54s\overline{4d}$,\\
& $3s^23p^63d\,^65s$,
$3s^23p^63d\,^6\overline{5d}$ \\
\hline
\noalign{\smallskip}
7-config &
$3s^23p^63d\,^6\overline{4d}$,
$3s^23p^63d\,^54s\overline{4d}$,
$3s3p^63d\,^74s$,
$3s3p^63d\,^8$ \\
\hline
\noalign{\smallskip}
BP new-TFDAc &
$3s^23p^63d\,^6\overline{4d}$,
$3s^23p^53d\,^64s4p$,
$3s^23p^63d\,^64p$,
$3s^23p^53d\,^74s$,
$3s^23p^53d\,^8$,
$3s^23p^43d\,^9$,\\
& $3s3p^63d\,^74s$,
$3s3p^63d\,^74p$,
$3s3p^63d\,^8$,
$3p^63d\,^84s$,
$3p^63d\,^74s4p$,
$3p^63d\,^9$,
$3p^63d\,^74s^2$ \\
\enddata
\tablecomments{Spectroscopic configurations give rise to the levels of interest and are common to all expansions.}
\end{deluxetable}

%%%%%%%%%%%%%%%%%%%%%%%%%%%%%%%%%%%%%%%%%%%%%%%%%%%%%%%%%%%%%%%%%%%%%%%%%%%%%%%

\begin{deluxetable}{llrrrrrrrr}
\tabletypesize{\scriptsize}
\tablecaption{Comparison of {\sc autostructure} energies with experiment \label{table:LSenergies}}
\tablewidth{0pt}
\tablehead{\colhead{Term} &\colhead{Expt} & \multicolumn{2}{c}{BP} & \multicolumn{2}{c}{BP exted TFDAc} & \multicolumn{2}{c}{Q96+4d$^2$-corr}  &  \multicolumn{2}{c}{NewTFDAc} \\
\cline{3-4}\cline{5-6}\cline{7-8}\cline{9-10}\\
  & & \colhead{$LS$}  &\colhead{$\langle JJ\rangle$} & \colhead{$LS$}  &\colhead{$\langle JJ\rangle$} & \colhead{$LS$} & \colhead{$\langle JJ\rangle$} & \colhead{$LS$} & \colhead{$\langle JJ\rangle$}
}
\startdata
$3d\,^64s\ ^6D$   &0.000 &0.000[01] &0.000[01] &0.000[01] &0.000[01] &0.000[01] &0.000[01] &0.001[02] &0.000[01] \\
$3d\,^7\ ^4F$     &0.018 &0.013[02] &0.034[02] &0.062[02] &0.077[02] &0.034[02] &0.058[02] &0.000[01] &0.019[02] \\
$3d\,^64s\ ^4D$   &0.072 &0.089[03] &0.093[03] &0.091[03] &0.090[03] &0.087[03] &0.093[03] &0.085[03] &0.084[03] \\
$3d\,^7\ ^4P$     &0.120 &0.124[04] &0.148[04] &0.164[04] &0.177[04] &0.158[04] &0.180[04] &0.110[04] &0.127[04] \\
$3d\,^7\ ^2G$     &0.143 &0.154[05] &0.179[05] &0.199[05] &0.214[07] &0.175[05] &0.199[05] &0.141[05] &0.160[05] \\
$3d\,^7\ ^2P$     &0.165 &0.167[06] &0.192[06] &0.212[07] &0.226[08] &0.208[06] &0.231[07] &0.155[06] &0.173[06] \\
$3d\,^7\ ^2H$     &0.184 &0.208[08] &0.234[10] &0.247[12] &0.271[14] &0.220[08] &0.244[10] &0.194[08] &0.214[08] \\
$3d\,^7\ ^2D$     &0.186 &0.187[07] &0.214[07] &0.232[09] &0.249[11] &0.230[10] &0.255[11] &0.174[07] &0.195[07] \\
$3d\,^64s\ ^4P$   &0.191 &0.229[11] &0.235[11] &0.224[08] &0.226[09] &0.224[09] &0.232[08] &0.230[10] &0.230[10] \\
$3d^64s\ ^4H$     &0.192 &0.208[09] &0.228[09] &0.211[06] &0.212[06] &0.212[07] &0.218[06] &0.224[09] &0.224[09] \\
$3d\,^64s\ ^4F$   &0.204 &0.291[15] &0.249[12] &0.232[10] &0.191[05] &0.234[11] &0.239[09] &0.245[12] &0.245[12] \\
$3d\,^54s^2\ ^6S$ &0.209 &0.214[10] &0.218[08] &0.243[11] &0.232[10] &0.425[16] &0.403[16] &0.236[11] &0.234[11] \\
$3d\,^64s\ ^4G$   &0.231 &0.275[12] &0.280[13] &0.256[13] &0.258[12] &0.260[12] &0.267[12] &0.276[14] &0.278[14] \\
$3d\,^64s\ ^2P$   &0.235 &0.281[14] &0.287[15] &0.277[15] &0.279[15] &0.275[14] &0.284[14] &0.280[15] &0.280[15] \\
$3d\,^64s\ 2H$    &0.235 &0.276[13] &0.283[14] &0.267[14] &0.266[13] &0.264[13] &0.273[13] &0.274[13] &0.273[13] \\
$3d\,^64s\ ^2F$   &0.246 &0.291[16] &0.298[16] &0.283[16] &0.284[16] &0.283[15] &0.290[15] &0.286[16] &0.290[16] \\
Core Energy       &      &$-2324.32$&$-2541.73$&$-2524.32$&$-2541.70$&$-2523.77$&$-2540.17$&$-2523.14$&$-2542.70$\\
\enddata
\tablecomments{Energies are given in Ryd. Numbers in square brackets indicate the relative term positions.}
\end{deluxetable}

%%%%%%%%%%%%%%%%%%%%%%%%%%%%%%%%%%%%%%%%%%%%%%%%%%%%%%%%%%%%%%%%%%%%%%%%%%%%%%%

\begin{deluxetable}{rllllrlll}
\tabletypesize{\scriptsize}
\tablecaption{Energy levels of the $3d\,^64s$, $3d\,^7$, and $3d\,^54s^2$ configurations \label{table:Elevels}}
\tablewidth{0pt}
\tablehead{\colhead{Index} & \colhead{Configuration} & \colhead{Level} & \colhead{Energy$^*$ (Ry)}
&&\colhead{Index} & \colhead{Configuration} & \colhead{Level} & \colhead{Energy$^*$ (Ry)}}
\startdata
1  & $3d\,^6(^5D)4s$ & $^6D_{9/2}$  & 0.0000000 && 27 & $3d\,^6(^3P2)4s$ &$^4P_{1/2}$  & 0.2042136 \\
2  & $3d\,^6(^5D)4s$ & $^6D_{7/2}$  & 0.0035065 && 28 & $3d\,^6(^3H)4s$  &$^4H_{13/2}$ & 0.1936589 \\
3  & $3d\,^6(^5D)4s$ & $^6D_{5/2}$  & 0.0060844 && 29 & $3d\,^6(^3H)4s$  &$^4H_{11/2}$ & 0.1952878 \\
4  & $3d\,^6(^5D)4s$ & $^6D_{3/2}$  & 0.0078607 && 30 & $3d\,^6(^3H)4s$  &$^4H_{9/2}$  & 0.1966664 \\
5  & $3d\,^6(^5D)4s$ & $^6D_{1/2}$  & 0.0089036 && 31 & $3d\,^6(^3H)4s$  &$^4H_{7/2}$  & 0.1978536 \\
6  & $3d\,^7$        & $^4F_{9/2}$  & 0.0170641 && 32 & $3d\,^6(^3F2)4s$ &$^4F_{9/2}$  & 0.2062854 \\
7  & $3d\,^7$        & $^4F_{7/2}$  & 0.0221447 && 33 & $3d\,^6(^3F2)4s$ &$^4F_{7/2}$  & 0.2078633 \\
8  & $3d\,^7$        & $^4F_{5/2}$  & 0.0258613 && 34 & $3d\,^6(^3F2)4s$ &$^4F_{5/2}$  & 0.2090388 \\
9  & $3d\,^7$        & $^4F_{3/2}$  & 0.0284084 && 35 & $3d\,^6(^3F2)4s$ &$^4F_{3/2}$  & 0.2098767 \\
10 & $3d\,^6(^5D)4s$ & $^4D_{7/2}$  & 0.0724940 && 36 & $3d\,^54s^2$     &$^6S_{5/2}$  & 0.2124859 \\
11 & $3d\,^6(^5D)4s$ & $^4D_{5/2}$  & 0.0764730 && 37 & $3d\,^6(^3G)4s$  &$^4G_{11/2}$ & 0.2317241 \\
12 & $3d\,^6(^5D)4s$ & $^4D_{3/2}$  & 0.0791021 && 38 & $3d\,^6(^3G)4s$  &$^4G_{9/2}$  & 0.2351555 \\
13 & $3d\,^6(^5D)4s$ & $^4D_{1/2}$  & 0.0806177 && 39 & $3d\,^6(^3G)4s$  &$^4G_{7/2}$  & 0.2367620 \\
14 & $3d\,^7$        & $^4P_{5/2}$  & 0.1227879 && 40 & $3d\,^6(^3G)4s$  &$^4G_{5/2}$  & 0.2374345 \\
15 & $3d\,^7$        & $^4P_{3/2}$  & 0.1245992 && 41 & $3d\,^6(^3P2)4s$ &$^2P_{3/2}$  & 0.2349939 \\
16 & $3d\,^7$        & $^4P_{1/2}$  & 0.1267101 && 42 & $3d\,^6(^3P2)4s$ &$^2P_{1/2}$  & 0.2454293 \\
17 & $3d\,^7$        & $^2G_{9/2}$  & 0.1443871 && 43 & $3d\,^6(^3H)4s$  &$^2H_{11/2}$ & 0.2384802 \\
18 & $3d\,^7$        & $^2G_{7/2}$  & 0.1491686 && 44 & $3d\,^6(^3H)4s$  &$^2H_{9/2}$  & 0.2401441 \\
19 & $3d\,^7$        & $^2P_{3/2}$  & 0.1673145 && 45 & $3d\,^6(^3F2)4s$ &$^2F_{7/2}$  & 0.2489119 \\
20 & $3d\,^7$        & $^2P_{1/2}$  & 0.1721090 && 46 & $3d\,^6(^3F2)4s$ &$^2F_{5/2}$  & 0.2516957 \\
21 & $3d\,^7$        &$^2H_{11/2}$  & 0.1853545 && 47 & $3d\,^6(^3G)4s$  &$^2G_{9/2}$  & 0.2769208 \\
22 & $3d\,^7$        & $^2H_{9/2}$  & 0.1895961 && 48 & $3d\,^6(^3G)4s$  &$^2G_{7/2}$  & 0.2803466 \\
23 & $3d\,^7$        & $^2D_{5/2}$  & 0.1869643 && 49 & $3d\,^6(^3D)4s$  &$^4D_{3/2}$  & 0.2858138 \\
24 & $3d\,^7$        & $^2D_{3/2}$  & 0.1941731 && 50 & $3d\,^6(^3D)4s$  &$^4D_{1/2}$  & 0.2858504 \\
25 & $3d\,^6(^3P2)4s$& $^4P_{5/2}$  & 0.1898222 && 51 & $3d\,^6(^3D)4s$  &$^4D_{5/2}$  & 0.2860280 \\
26 & $3d\,^6(^3P2)4s$& $^4P_{3/2}$  & 0.1987661 && 52 & $3d\,^6(^3D)4s$  &$^4D_{7/2}$  & 0.2868958 \\
\enddata
\tablecomments{$^*$ From \citet{NIST}}
\end{deluxetable}

%%%%%%%%%%%%%%%%%%%%%%%%%%%%%%%%%%%%%%%%%%%%%%%%%%%%%%%%%%%%%%%%%%%%%%%%%%%%%%%

\begin{deluxetable}{cllllllllll}
\tabletypesize{\scriptsize}
\tablecaption{Theoretical radiative widths (s$^{-1}$) for even-parity levels \label{table:lifetimes}}
\tablewidth{0pt}
\tablehead{\colhead{Level} &\colhead{SST$^a$}&\colhead{HFR$^b$}&\colhead{HFR$^c$}&\colhead{CIV3$^d$}&\colhead{TFDAc$^e$}&\colhead{Q96$^f$}&
\colhead{7-config$^g$}&\colhead{TFDAc$^h$}&\colhead{Rec$^i$}&\colhead{Unc$^j$}}
%\tablehead{
%\colhead{Level}  &\colhead{SST(QDZ96} & \colhead{HFR(QDZ96)} & \colhead{HFR new} & \colhead{CIV3(DH11)} &
%\colhead{BP Extend. TFDAc} & \colhead{Q96+4d$^2$-corr}  & \colhead{7-config} & \colhead{NewTFDAc}
%&   & \colhead{Recom} & \colhead{$\Delta(1/\tau)$ (\%)} }
\startdata
2   &2.13$-$3&2.05$-$3& 2.14$-$3& 2.14$-$3&2.13$-$3&2.13$-$3&2.13$-$3&2.13$-$3&2.12$-$3& 1.31 \\
3   &1.57$-$3&1.56$-$3& 1.58$-$3& 1.58$-$3&1.57$-$3&1.57$-$3&1.57$-$3&1.57$-$3&1.57$-$3& 0.38 \\
4   &7.19$-$4&7.28$-$4& 7.21$-$4& 7.21$-$4&7.18$-$4&7.18$-$4&7.18$-$4&7.18$-$4&7.20$-$4& 0.46 \\
5   &1.89$-$4&1.94$-$4& 1.93$-$4& 1.89$-$4&1.88$-$4&1.88$-$4&1.88$-$4&1.88$-$4&1.89$-$4& 1.10 \\
6   &9.99$-$5&5.83$-$5& 6.49$-$5& 1.42$-$4&8.08$-$6&1.92$-$5&1.62$-$5&6.78$-$5&6.8 $-$5& 30   \\
7   &5.95$-$3&6.07$-$3& 5.94$-$3& 6.01$-$3&5.84$-$3&5.85$-$3&5.85$-$3&5.90$-$3&5.92$-$3& 1.32 \\
8   &3.99$-$3&4.21$-$3& 3.99$-$3& 4.03$-$3&3.92$-$3&3.93$-$3&3.92$-$3&3.95$-$3&3.99$-$3& 2.29 \\
9   &1.44$-$3&1.55$-$3& 1.44$-$3& 1.46$-$3&1.41$-$3&1.42$-$3&1.42$-$3&1.43$-$3&1.44$-$3& 2.92 \\
10  &1.43$-$2&1.46$-$2& 1.43$-$2& 1.40$-$2&1.79$-$2&1.77$-$2&1.22$-$2&1.39$-$2&1.48$-$2& 12.6 \\
11  &1.38$-$2&1.40$-$2& 1.28$-$2& 1.30$-$2&1.67$-$2&1.48$-$2&1.19$-$2&1.35$-$2&1.38$-$2& 9.90 \\
12  &1.24$-$2&1.27$-$2& 1.19$-$2& 1.16$-$2&1.55$-$2&1.36$-$2&1.07$-$2&1.23$-$2&1.26$-$2& 10.8 \\
13  &1.49$-$2&1.55$-$2& 1.08$-$2& 1.08$-$2&1.53$-$2&1.31$-$2&1.01$-$2&1.16$-$2&1.26$-$2& 16.8 \\
14  &5.65$-$2&5.15$-$2& 4.72$-$2& 5.24$-$2&5.10$-$2&4.32$-$2&4.13$-$2&5.38$-$2&5.01$-$2& 10.2 \\
15  &5.06$-$2&4.73$-$2& 4.27$-$2& 4.76$-$2&4.86$-$2&4.12$-$2&3.97$-$2&4.53$-$2&4.54$-$2& 7.92 \\
16  &5.04$-$2&4.76$-$2& 4.27$-$2& 4.76$-$2&4.98$-$2&4.15$-$2&3.97$-$2&4.48$-$2&4.54$-$2& 8.16 \\
17  &1.94$-$1&2.02$-$1& 1.93$-$1& 2.52$-$1&2.59$-$1&1.42$-$1&1.36$-$1&2.45$-$1&2.08$-$1& 22.4 \\
18  &1.05$-$1&1.11$-$1& 1.06$-$1& 1.38$-$1&1.39$-$1&7.69$-$2&7.36$-$2&1.32$-$1&1.13$-$1& 22.2 \\
19  &1.64$-$1&1.59$-$1& 1.60$-$1& 2.20$-$1&2.52$-$1&1.19$-$1&1.33$-$1&2.70$-$1&1.94$-$1& 30.5 \\
20  &9.58$-$2&9.43$-$2& 1.00$-$1& 1.27$-$1&1.34$-$1&7.55$-$2&8.55$-$2&1.42$-$1&1.11$-$1& 23.1 \\
21  &1.48$-$2&1.58$-$2& 1.60$-$2& 1.95$-$2&1.20$-$1&1.25$-$2&1.21$-$2&1.45$-$2&1.50$-$2& 15.4 \\
22  &6.02$-$2&6.42$-$2& 6.44$-$2& 7.84$-$2&3.16$-$2&4.69$-$2&4.33$-$2&9.95$-$2&6.07$-$2& 32.9 \\
23  &3.78$-$1&3.77$-$1& 4.07$-$1& 5.30$-$1&5.91$-$1&2.55$-$1&2.35$-$1&5.73$-$1&4.35$-$1& 31.6 \\
24  &5.03$-$1&5.15$-$1& 4.94$-$1& 6.69$-$1&7.53$-$1&3.62$-$1&3.34$-$1&7.25$-$1&5.65$-$1& 28.3 \\
25  &1.09$+$0&1.20$+$0& 1.11$+$0& 1.17$+$0&1.49$+$0&1.15$+$0&1.24$+$0&9.92$-$1&1.16$+$0& 13.0 \\
26  &1.30$+$0&1.44$+$0& 1.41$+$0& 1.45$+$0&1.72$+$0&1.37$+$0&1.43$+$0&1.56$+$0&1.43$+$0& 10.6 \\
27  &1.40$+$0&1.56$+$0& 1.53$+$0& 1.60$+$0&1.87$+$0&1.47$+$0&1.57$+$0&1.28$+$0&1.51$+$0& 12.0 \\
28  &5.56$-$1&6.05$-$1& 5.40$-$1& 4.76$-$1&6.67$-$1&4.77$-$1&4.85$-$1&5.43$-$1&5.45$-$1& 12.4 \\
29  &5.21$-$1&5.60$-$1& 5.05$-$1& 4.47$-$1&8.00$-$1&4.50$-$1&4.58$-$1&5.16$-$1&5.30$-$1& 20.4 \\
30  &4.93$-$1&5.25$-$1& 4.78$-$1& 4.30$-$1&6.37$-$1&4.28$-$1&4.36$-$1&4.86$-$1&4.89$-$1& 13.2 \\
31  &4.68$-$1&4.94$-$1& 4.52$-$1& 4.02$-$1&1.24$+$0&4.06$-$1&4.13$-$1&4.62$-$1&4.45$-$1& 7.58 \\
32  &1.03$+$0&1.14$+$0& 1.13$+$0& 1.21$+$0&1.20$+$0&1.08$+$0&1.05$+$0&9.54$-$1&1.08$+$0& 8.79 \\
33  &8.77$-$1&9.76$-$1& 9.63$-$1& 1.01$+$0&1.03$+$0&9.06$-$1&8.95$-$1&8.22$-$1&9.22$-$1& 8.29 \\
34  &7.15$-$1&7.98$-$1& 7.81$-$1& 8.08$-$1&8.59$-$1&7.17$-$1&7.26$-$1&6.83$-$1&7.52$-$1& 8.21 \\
35  &5.76$-$1&6.44$-$1& 6.15$-$1& 6.21$-$1&7.06$-$1&5.46$-$1&5.76$-$1&5.60$-$1&6.01$-$1& 8.54 \\
36  &3.81$+$0&4.54$+$0& 4.98$+$0& 4.29$+$0&5.57$+$0&4.68$+$0&4.86$+$0&2.30$+$0&4.15$+$0& 27.8 \\
37  &1.29$+$0&1.42$+$0& 1.29$+$0& 1.04$+$0&1.32$+$0&6.73$-$1&1.23$+$0&7.98$-$1&1.10$+$0& 25.2 \\
38  &1.32$+$0&1.44$+$0& 1.34$+$0& 1.16$+$0&1.54$+$0&7.60$-$1&1.24$+$0&8.97$-$1&1.18$+$0& 22.9 \\
39  &1.30$+$0&1.41$+$0& 1.33$+$0& 1.16$+$0&1.61$+$0&1.16$+$0&1.21$+$0&1.29$+$0&1.31$+$0& 10.7 \\
40  &1.58$+$0&1.37$+$0& 1.30$+$0& 1.14$+$0&1.59$+$0&1.15$+$0&1.20$+$0&1.26$+$0&1.32$+$0& 12.8 \\
41  &5.14$-$1&5.33$-$1& 5.71$-$1& 6.83$-$1&5.32$-$1&6.19$-$1&5.32$-$1&4.15$-$1&5.35$-$1& 16.1 \\
42  &6.30$-$1&6.67$-$1& 7.06$-$1& 8.33$-$1&6.44$-$1&7.36$-$1&6.75$-$1&5.17$-$1&6.58$-$1& 15.2 \\
43  &1.52$-$1&1.92$-$1& 1.96$-$1& 2.69$-$1&1.31$-$1&7.12$-$1&1.01$-$1&7.29$-$1&1.74$-$1& 34.1 \\
44  &6.61$-$2&8.23$-$2& 8.23$-$2& 8.65$-$2&3.83$+$0&5.62$-$1&5.99$-$2&7.31$-$1&7.54$-$2& 15.5 \\
45  &5.21$-$1&5.41$-$1& 5.91$-$1& 6.57$-$1&5.90$-$1&6.16$-$1&5.67$-$1&4.32$-$1&5.64$-$1& 12.1 \\
46  &3.28$-$1&3.41$-$1& 3.75$-$1& 3.95$-$1&3.63$-$1&3.90$-$1&3.41$-$1&2.86$-$1&3.45$-$1& 11.8 \\
47  &2.98$-$1&3.20$-$1& 3.00$-$1& 2.86$-$1&5.18$-$1&3.00$-$1&2.81$-$1&2.80$-$1&2.92$-$1& 5.19 \\
48  &2.75$-$1&2.92$-$1& 2.75$-$1& 2.76$-$1&3.49$-$1&2.68$-$1&2.58$-$1&2.82$-$1&2.84$-$1& 9.19 \\
49  &1.60$+$0&1.81$+$0& 1.73$+$0& 1.57$+$0&2.04$+$0&1.57$+$0&1.59$+$0&1.54$+$0&1.67$+$0& 9.95 \\
50  &1.62$+$0&1.82$+$0& 1.76$+$0& 1.60$+$0&2.04$+$0&1.61$+$0&1.61$+$0&1.57$+$0&1.69$+$0& 9.34 \\
51  &1.62$+$0&1.83$+$0& 1.75$+$0& 1.58$+$0&2.05$+$0&1.58$+$0&1.60$+$0&1.59$+$0&1.69$+$0& 9.6  \\
52  &1.76$+$0&1.99$+$0& 1.91$+$0& 1.73$+$0&2.21$+$0&1.71$+$0&1.74$+$0&1.72$+$0&1.83$+$0& 9.41 \\
\enddata
\tablecomments{The level radiative width is defined as $A_i = \sum_j A(i\to j)$. The last two columns present our recommended values and their uncertainties. $a\pm b\equiv a\times 10^{\pm b}$.}
\tablenotetext{a}{SST(QDZ96)}
\tablenotetext{b}{HFR(QDZ96)}
\tablenotetext{c}{HFR new}
\tablenotetext{d}{CIV3(DH11)}
\tablenotetext{e}{BP extend TFDAc}
\tablenotetext{f}{Q96+4d$^2$-corr}
\tablenotetext{g}{7-config}
\tablenotetext{h}{NewTFDAc}
\tablenotetext{i}{Recommended value}
\tablenotetext{j}{Uncertainty (\%)}
\end{deluxetable}

%%%%%%%%%%%%%%%%%%%%%%%%%%%%%%%%%%%%%%%%%%%%%%%%%%%%%%%%%%%%%%%%%%%%%%%%%%%%%%%

\begin{deluxetable}{rrlllllllllr}
\tabletypesize{\scriptsize}
\tablecaption{Theoretical branching ratios for forbidden transitions among even parity levels \label{table:branching}}
\tablewidth{0pt}
\tablehead{\colhead{$i$}&\colhead{$j$}&\colhead{SST$^a$}&\colhead{HFR$^b$}&\colhead{HFR$^c$}&\colhead{CIV3$^d$}&
\colhead{TFDAc$^e$}&\colhead{Q96$^f$}&\colhead{7-config$^g$}&\colhead{TFDAc$^h$}&\colhead{Rec$^i$}&\colhead{Unc$^j$}}
\startdata
2  &1& 1.00$+$0&1.00$+$0&1.00$+$0&1.00$+$0&1.00$+$0&1.00$+$0&1.00$+$0&1.00$+$0&1.00$+$0&0.00\\
3  &2& 1.00$+$0&1.00$+$0&1.00$+$0&1.00$+$0&1.00$+$0&1.00$+$0&1.00$+$0&1.00$+$0&1.00$+$0&0.00\\
4  &3& 1.00$+$0&1.00$+$0&1.00$+$0&1.00$+$0&1.00$+$0&1.00$+$0&1.00$+$0&1.00$+$0&1.00$+$0&0.00\\
5  &4& 1.00$+$0&1.00$+$0&1.00$+$0&1.00$+$0&1.00$+$0&1.00$+$0&1.00$+$0&1.00$+$0&1.00$+$0&0.00\\
6  &1& 9.16$-$1&9.18$-$1&9.21$-$1&9.15$-$1&9.29$-$1&9.32$-$1&9.32$-$1&9.06$-$1&9.21$-$1&1.01\\
6  &2& 8.37$-$2&8.18$-$2&7.80$-$2&8.17$-$2&7.00$-$2&6.72$-$2&6.79$-$2&9.40$-$2&7.80$-$2&11.9\\
7  &6& 9.82$-$1&9.90$-$1&9.88$-$1&9.75$-$1&9.98$-$1&9.97$-$1&9.97$-$1&9.88$-$1&9.89$-$1&0.82\\
8  &7& 9.82$-$1&9.90$-$1&9.87$-$1&9.75$-$1&9.97$-$1&9.95$-$1&9.97$-$1&9.90$-$1&9.89$-$1&0.78\\
9  &8& 9.79$-$1&9.87$-$1&9.86$-$1&9.73$-$1&1.00$+$0&9.93$-$1&9.93$-$1&9.86$-$1&9.87$-$1&0.86\\
10 &1& 3.31$-$1&3.55$-$1&3.95$-$1&3.76$-$1&2.55$-$1&4.62$-$1&3.59$-$1&3.49$-$1&3.60$-$1&16.3\\
10 &2& 9.16$-$2&9.86$-$2&1.10$-$1&1.06$-$1&7.71$-$2&1.36$-$1&1.12$-$1&1.03$-$1&1.04$-$1&16.4\\
10 &3& 5.89$-$2&6.12$-$2&6.88$-$2&6.94$-$2&4.68$-$2&8.64$-$2&6.78$-$2&6.15$-$2&6.51$-$2&17.4\\
10 &6& 4.18$-$1&3.92$-$1&3.44$-$1&3.62$-$1&4.84$-$1&2.53$-$1&3.72$-$1&3.90$-$1&3.77$-$1&17.5\\
10 &7& 9.23$-$2&8.56$-$2&7.55$-$2&8.00$-$2&1.20$-$1&5.47$-$2&8.28$-$2&8.63$-$2&8.47$-$2&21.5\\
\enddata
\tablecomments{The branching ratio is given by $b(i\to j) = A(i\to j)\times \tau(i) = A(i\to j)/\sum_k A(i\to k)$. The last
two columns tabulate our recommended values and their uncertainties. $a\pm b\equiv a\times 10^{\pm b}$.}
\tablenotetext{a}{SST(QDZ96)}
\tablenotetext{b}{HFR(QDZ96)}
\tablenotetext{c}{HFR new}
\tablenotetext{d}{CIV3(DH11)}
\tablenotetext{e}{BP extend TFDAc}
\tablenotetext{f}{Q96+4d$^2$-corr}
\tablenotetext{g}{7-config}
\tablenotetext{h}{NewTFDAc}
\tablenotetext{i}{Recommended value}
\tablenotetext{j}{Uncertainty (\%)}
\end{deluxetable}

%%%%%%%%%%%%%%%%%%%%%%%%%%%%%%%%%%%%%%%%%%%%%%%%%%%%%%%%%%%%%%%%%%%%%%%%%%%%%%%

\begin{deluxetable}{cllllllllll}
\tabletypesize{\scriptsize}
\tablecaption{Comparison of theoretical lifetimes (s$^{-1}$) with experiment \label{table:lifetcomp}}
\tablewidth{0pt}
\tablehead{\colhead{Level} &\colhead{SST$^a$}&\colhead{HFR$^b$}&\colhead{HFR$^c$}&\colhead{CIV3$^d$}&\colhead{TFDAc$^e$}&\colhead{Q96$^f$}
        &\colhead{7-config$^g$}&\colhead{TFDAc$^h$}& \colhead{Rec$^i$}&\colhead{Expt$^j$}}
\startdata
36& 0.262& 0.220& 0.201& 0.233& 0.180& 0.214& 0.206& 0.435 & $0.24\pm 0.07$ & $0.23\pm 0.03$\\ 37& 0.775& 0.704& 0.775& 0.962& 0.758& 1.49 & 0.813& 1.25  & $0.91\pm 0.23$ & $0.75\pm 0.01$\\
38& 0.758& 0.694& 0.746& 0.862& 0.649& 1.32 & 0.806& 1.11  & $0.85\pm 0.20$ & $0.65\pm 0.02$\\	
43&  6.58& 5.21 & 5.10 & 3.72 & 7.63 & 1.40 & 9.90 & 1.37  & $5.8\pm  2.0 $ & $3.8\pm  0.3$ \\
\enddata
\tablenotetext{a}{SST(QDZ96)}
\tablenotetext{b}{HFR(QDZ96)}
\tablenotetext{c}{HFR new}
\tablenotetext{d}{CIV3(DH11)}
\tablenotetext{e}{BP extend TFDAc}
\tablenotetext{f}{Q96+4d$^2$-corr}
\tablenotetext{g}{7-config}
\tablenotetext{h}{NewTFDAc}
\tablenotetext{i}{Recommended theoretical lifetimes}
\tablenotetext{j}{Experimental lifetimes from the Ferrum Project \citep{har03, gur09}}
\end{deluxetable}

%%%%%%%%%%%%%%%%%%%%%%%%%%%%%%%%%%%%%%%%%%%%%%%%%%%%%%%%%%%%%%%%%%%%%%%%%%%%%%%

\begin{deluxetable}{ll}
\tabletypesize{\scriptsize}
\tablecaption{Comparison of theoretical and observed branching ratios \label{table:chis}}
\tablewidth{0pt}
\tablehead{
\colhead{Calculation} & \colhead{$\chi^2$ }}
\startdata
{SST(QDZ96)}      & 5.83   \\
{HFR(QDZ96)}      & 5.09   \\
{HFR new}         & 5.16   \\
{CIV3(DH11)}      & 4.76   \\
{BP extend TFDAc}         & 358    \\
{Q96+4d$^2$-corr} & 12.64  \\
{7-config}         & 4.10   \\
{NewTFDAc}        & 4.92   \\
{Recom.($\delta R_{\rm th}=0$)}$^a$ & 4.34 \\
{Recommended}$^b$     & 1.38 \\
\enddata
\tablecomments{The comparison (in terms of the reduced $\chi^2$) is for 106 observed branching ratios with common upper levels.}
\tablenotetext{a}{Excludes estimated theoretical uncertainties.}
\tablenotetext{b}{Includes estimated theoretical uncertainties.}
\end{deluxetable}

%%%%%%%%%%%%%%%%%%%%%%%%%%%%%%%%%%%%%%%%%%%%%%%%%%%%%%%%%%%%%%%%%%%%%%%%%%%%%%%

\begin{deluxetable}{rrrrrrrrrrrrr}
\tabletypesize{\scriptsize}
\tablecaption{Effective collision strengths from the ground level of [\ion{Fe}{2}] at $10^4$~K \label{table:upsil104}}
\tablewidth{0pt}
\tablehead{\colhead{j} &\colhead{Q+RM$^a$} &\colhead{Q+RM$^b$} &\colhead{Q+RM$^c$} &\colhead{Q+RM$^d$}
 &\colhead{7-config$^e$} &\colhead{DARC$^f$} &\colhead{Mean$^g$} &\colhead{$\sigma^h$} &\colhead{ZP96$^i$} &\colhead{BP96$^k$}
 &\colhead{RH07$^k$}   }
\startdata
 2& 1.81$+$0& 2.05$+$0& 2.80$+$0& 2.24$+$0 &2.85$+$0 & 5.16$+$0& 2.31$+$0 &16.0 &5.52$+$0 &4.65$+$0 &4.84$+$0\\
 3& 3.36$-$1& 3.89$-$1& 6.52$-$1& 4.93$-$1 &4.20$-$1 & 1.20$+$0& 4.21$-$1 &23.8 &1.49$+$0 &1.29$+$0 &1.12$+$0\\
 4& 1.54$-$1& 1.58$-$1& 3.58$-$1& 2.33$-$1 &1.78$-$1 & 5.11$-$1& 1.95$-$1 &35.3 &6.84$-$1 &8.13$-$1 &5.29$-$1\\
 5& 7.16$-$2& 7.33$-$2& 1.81$-$1& 1.16$-$1 &8.57$-$2 & 2.22$-$1& 9.51$-$2 &37.8 &2.84$-$1 &4.33$-$1 &2.47$-$1\\
 6& 2.05$+$0& 1.99$+$0& 2.56$+$0& 2.20$+$0 &1.84$+$0 & 3.07$+$0& 2.05$+$0 &11.5 &3.60$+$0 &1.31$+$0 &2.83$+$0\\
 7& 7.46$-$1& 7.53$-$1& 9.27$-$1& 8.74$-$1 &6.78$-$1 & 1.33$+$0& 7.64$-$1 &11.6 &1.51$+$0 &6.14$-$1 &1.21$+$0\\
 8& 1.57$-$1& 1.59$-$1& 1.90$-$1& 2.26$-$1 &1.42$-$1 & 4.08$-$1& 1.64$-$1 &18.1 &4.97$-$1 &1.35$-$1 &2.98$-$1\\
 9& 3.18$-$2& 2.72$-$2& 3.29$-$2& 6.63$-$2 &2.25$-$2 & 1.01$-$2& 3.09$-$2 &48.1 &1.37$-$1 &1.65$-$2 &5.84$-$2\\
10& 1.61$+$0& 1.59$+$0& 2.02$+$0& 1.78$+$0 &4.26$+$0 & 9.04$+$0& 2.37$+$0 &45.5 &1.10$+$1 &1.43$+$1 &1.04$+$1\\
11& 1.18$-$1& 1.02$-$1& 1.47$-$1& 1.51$-$1 &1.17$-$1 & 6.22$-$1& 1.11$-$1 &24.2 &5.59$-$1 &5.72$-$1 &5.26$-$1\\
12& 5.63$-$2& 4.45$-$2& 6.92$-$2& 8.79$-$2 &4.27$-$2 & 2.14$-$1& 5.13$-$2 &34.1 &1.91$-$1 &3.86$-$1 &1.79$-$1\\
13& 2.21$-$2& 2.19$-$2& 3.01$-$2& 3.89$-$2 &1.94$-$2 & 7.61$-$2& 2.33$-$2 &31.0 &6.01$-$2 &2.33$-$2 &6.49$-$2\\
14& 7.53$-$1& 6.40$-$1& 8.90$-$1& 7.30$-$1 &6.75$-$1 & 9.42$-$1& 7.02$-$1 &11.0 &9.48$-$1 &5.42$-$1 &9.47$-$1\\
15& 3.69$-$1& 3.78$-$1& 4.78$-$1& 3.91$-$1 &4.06$-$1 & 4.80$-$1& 3.98$-$1 & 8.3 &5.02$-$1 &3.19$-$1 &5.22$-$1\\
16& 1.19$-$2& 7.60$-$3& 9.28$-$3& 9.40$-$3 &5.46$-$3 & 5.89$-$2& 7.44$-$3 &21.8 &3.28$-$2 &8.20$-$3 &6.11$-$2\\
17& 2.35$-$1& 7.73$-$2& 1.72$-$1& 2.01$-$1 &3.42$-$2 & 9.79$-$2& 1.08$-$1 &59.1 &  --     & --      &9.73$-$2\\
18& 1.24$-$1& 4.35$-$2& 6.24$-$2& 1.12$-$1 &1.28$-$2 & 5.65$-$2& 5.19$-$2 &65.4 &  --     & --      &6.25$-$2\\
19& 2.84$-$2& 1.78$-$2& 2.76$-$2& 3.04$-$2 &8.05$-$3 & 2.98$-$2& 1.77$-$2 &55.5 &  --     & --      &4.54$-$2\\
20& 4.52$-$3& 7.12$-$5& 3.22$-$3& 6.95$-$3 &3.70$-$4 & 8.04$-$3& 1.84$-$2 &244. &  --     & --      &1.32$-$2\\
21& 8.65$-$2& 3.39$-$5& 1.18$-$1& 1.02$-$1 &4.08$-$3 & 4.46$-$1& 4.23$-$2 &108. &  --     & --      &1.55$-$1\\
22& 6.33$-$2& 1.53$-$3& 5.04$-$2& 7.77$-$2 &1.79$-$3 & 1.74$-$1& 2.59$-$2 &107. &  --     & --      &3.90$-$2\\
23& 3.86$-$2& 3.02$-$1& 3.12$-$2& 5.29$-$2 &3.26$-$3 & 4.60$-$2& 8.80$-$2 &114. &  --     & --      &4.66$-$2\\
24& 9.19$-$3& 9.82$-$2& 6.72$-$3& 1.50$-$2 &8.14$-$4 & 9.25$-$3& 2.27$-$2 &1430 &  --     & --      &1.63$-$2\\
25& 2.30$-$1& 1.23$-$1& 3.35$-$1& 2.85$-$1 &1.44$-$1 & 3.89$-$1& 1.85$-$1 &56.0 &3.08$-$1 &2.15$-$1 &2.67$-$1\\
26& 5.85$-$2& 4.14$-$2& 8.79$-$2& 6.86$-$2 &3.49$-$2 & 1.02$-$1& 6.48$-$2 &53.7 &9.92$-$2 &5.17$-$2 &7.01$-$2\\
27& 1.78$-$3& 2.26$-$4& 2.62$-$3& 2.14$-$3 &9.46$-$4 & 1.69$-$2& 3.67$-$3 &172. &8.00$-$3 &4.30$-$3 &6.12$-$3\\
28& 5.42$-$1& 5.66$-$1& 8.14$-$1& 6.88$-$1 &3.26$-$1 & 7.97$-$1& 5.44$-$1 &30.2 &6.31$-1$ &1.96$-$1 &7.19$-$1\\
29& 3.13$-$1& 3.45$-$1& 4.57$-$1& 4.50$-$1 &1.77$-$1 & 2.70$-$2& 3.14$-$1 &33.6 &3.11$-$1 &9.55$-$2 &3.74$-$1\\
30& 1.18$-$1& 1.39$-$1& 1.71$-$1& 1.88$-$1 &6.60$-$2 & 6.76$-$1& 1.19$-$1 &37.5 &9.51$-$2 &2.76$-$2 &1.46$-$1\\
\enddata
\tablecomments{$a\pm b\equiv a\times 10^{\pm b}$.}
\tablenotetext{a}{Q+RM-ns}
\tablenotetext{b}{Q+RM-shift}
\tablenotetext{c}{Q+RM-ns-RA=8}
\tablenotetext{d}{Q+RM-ns-RA=14.5}
\tablenotetext{e}{7-config}
\tablenotetext{f}{DARC} 
\tablenotetext{g}{Mean of effective collision strengths obtained in the present calculations}
\tablenotetext{h}{Standard deviation (\%) for effective collision strengths obtained in the present calculations}
\tablenotetext{i}{\citet{zhang95}}
\tablenotetext{j}{\citet{bau96}}
\tablenotetext{k}{\citet{ram07}}
\end{deluxetable}

%%%%%%%%%%%%%%%%%%%%%%%%%%%%%%%%%%%%%%%%%%%%%%%%%%%%%%%%%%%%%%%%%%%%%%%%%%%%%%%

\begin{deluxetable}{lccc}
\tabletypesize{\scriptsize}
\tablecaption{Comparison of normalized line intensities in HH~202 with model predictions \label{table:linecomps}}
\tablewidth{0pt}
\tablehead{
\colhead{Model}  &\colhead{$\chi^2$} & \colhead{$T_e$} & \colhead{$n_e$}\\
                 &                   & \colhead{($10^4$~K)}      & \colhead{($10^4$~cm$^{-3}$)}}
\startdata
ZP96$^a$ & 1.40 & 0.9 & 4.7 \cr
BP98$^b$ & 1.39 & 1.0 & 4.5 \cr
RH07$^c$ & 1.60 & 1.2 & 3.0 \cr
Pres$^d$ & 3.13 & 1.0 & 2.0 \cr
Pres$^e$ & 1.28 & 1.2 & 10. \cr
Pres$^f$ & 1.05 & 1.0 & 7.0 \cr
Pres$^g$ & 1.01 & 1.15& 6.6 \cr
\enddata
\tablecomments{Each model uses a specific set of effective collision strengths. The listed reduced $\chi^2$ corresponds to the best match with observation at the quoted temperature and density.}
\tablenotetext{a}{Collision strengths from \citet{zhang95}.}
\tablenotetext{b}{Collision strengths from \citet{baup98}.}
\tablenotetext{c}{Collision strengths from \citet{ram07} and \citet{baup98}.}
\tablenotetext{d}{Present {\sc darc} calculation.}
\tablenotetext{e}{Present {\sc bprm} calculation with 114 levels.}
\tablenotetext{f}{Present {\sc bprm} calculation with 63 levels.}
\tablenotetext{g}{Present {\sc bprm} calculation with 7-config.}
\end{deluxetable}

%%%%%%%%%%%%%%%%%%%%%%%%%%%%%%%%%%%%%%%%%%%%%%%%%%%%%%%%%%%%%%%%%%%%%%%%%%%%%%%

\end{document}